\documentclass[twocolumn]{aastex6}
\fullcollaborationName{The Friends of AASTeX Collaboration}

\begin{document}


\title{Perpendicular diffusion of solar energetic particles; model results and implications for electrons}


\author{R Du Toit Strauss\altaffilmark{1}}
\affil{Center for Space Research, North-West University, Potchefstroom, 2522, South Africa}
\affil{National Institute for Theoretical Physics (NITheP), Gauteng, South Africa}

\author{Nina Dresing}
\affil{Institut f{\"u}r Experimentelle und Angewandte Physik, Christian-Albrechts-Universit{\"a}t zu Kiel, Germany}

\and

\author{N Eugene Engelbrecht}
\affil{Center for Space Research, North-West University, Potchefstroom, 2522, South Africa}


\altaffiltext{1}{dutoit.strauss@nwu.ac.za}

\begin{abstract}

The processes responsible for the effective longitudinal transport of solar energetic particles (SEPs) are still not completely understood. We address this issue by simulating SEP electron propagation using a spatially 2D transport model that includes perpendicular diffusion. By implementing, as far as possible, the most reasonable estimates of the transport (diffusion) coefficients, we compare our results, in a qualitative manner, to recent observations {at energies of 55 -- 105 keV}, focusing on the longitudinal distribution of the peak intensity, the maximum anisotropy and the onset time. By using transport coefficients which are derived from first principles, we limit the number of free parameters in the model to: (i) the probability of SEPs following diffusing magnetic field lines, quantified by $a \in [0,1]$, and (ii) the broadness of the Gaussian injection function. It is found that the model solutions are extremely sensitive to the magnitude of the {perpendicular} diffusion coefficient and relatively insensitive to the form of the injection function as long as a reasonable value of $a=0.2$ is used. We illustrate the effects of perpendicular diffusion on the model solutions and discuss the viability of this process as a dominant mechanism by which SEPs are transported in longitude. Lastly, we try to quantity the effectiveness of perpendicular diffusion as an interplay between the magnitude of the relevant diffusion coefficient and the SEP intensity gradient driving the diffusion process. It follows that perpendicular diffusion is extremely effective early in a SEP event when large intensity gradients are present, while the effectiveness quickly decreases with time thereafter.

\end{abstract}




\section{Introduction}

{Transient solar phenomena are believed to be responsible for the acceleration of solar energetic particles (SEPs), which are usually grouped into two distinct classifications \citep{remaes1999} according to where and how they were accelerated: smaller, short lived, impulsive electron-rich events are believed to be accelerated close to the Sun and are generally associated with solar flares, while the larger, so-called gradual proton-rich events, are associated with coronal mass ejections \citep[see the review by][]{remaesreview}. However, recent studies have shown that this classification might be too simple: The unexpected observations of widespread $^3$He and electron events \citep[e.g.][]{widenbecketal2013, dresingetal2012, dresing2014} suggest that further mechanisms may play a role, which were not taken into account by the old classification. These mechanisms are among others perpendicular transport close to the Sun or in the interplanetary medium, and acceleration of $^3$He and electrons in shocks. In this study we assume an impulsive acceleration of electrons, occurring in a compact region close to the Sun,} through, for instance, magnetic reconnection in the solar flares themselves or through shock acceleration occurring at coronal shocks. After being accelerated to relativistic energies, these SEPs are released and propagate along the interplanetary magnetic field to reach the near Earth environment, where they are observed {\it in situ} by a fleet of spacecraft. We focus on the transport of $\sim 85$ keV electrons, which are, from a modeling point-of-view, ideal test-particles as they suffer little, if any, re-acceleration by traveling shocks {\citep[e.g.][]{dresing2016}}, {adiabatic energy losses are usually negligible for relativistic SEPs \citep[e.g.][]{ruffolo1995},} and due to their high {propagation speeds, drifts (including co-rotation) are also usually negligible.} These assumptions allow us to simplify the modeling approach and focus on a specific topic, {the main topic of investigation in this paper being the longitudinal spread of (impulsively accelerated) energetic electron events. For some of these events, the so-called widespread events, SEPs are observed to cover up to $360^{\circ}$ in longitude at Earth \citep[e.g.][]{widenbecketal2013, dresingetal2012, dresing2014}.} Such an unexpectedly broad distribution can be due to either a broad injection region, such as an extended source \citep[due to various processes; see e.g.][]{cliveretal1995,kleinetal2008,larioetal2016}, or due to effective diffusion perpendicular to the mean field, or, of course, a combination of these processes. In this paper, we vary the broadness of the source region and the effectiveness of perpendicular diffusion to examine the effects thereof on simulated intensities and compare the model results, in a qualitative fashion, to observations. {Perpendicular diffusion has been included in previous transport models \citep[e.g.][]{zhangetal2009,wolfgang,drogeteal2014,he2015}, and these authors have been able to explain the observed broadness of the wide-spread events in terms of cross-field diffusion. A major criticism of these models is, however, that they treat the diffusion coefficients (especially these governing perpendicular diffusion) as adjustable parameters that are tuned in an {\it ad-hoc} fashion to reproduce SEP observations without any theoretical motivation \citep[see e.g.][]{remaesreview2}. In contrast to most previous modeling studies, we therefore try,} as far as possible, to implement theoretically derived, {i.e. derived from first principles and not prescribed in an {\it ad-hoc} manner,} transport coefficients in order to limit the number of free parameters in the model and, in so doing, move away from the phenomenological method of prescribing perpendicular diffusion, towards a self-consistent description of SEP transport.

\section{Numerical transport model}

The propagation of energetic electrons is described by the so-called focussed transport equation \citep[e.g.][]{skilling1971}, given by 

\begin{eqnarray}
\label{Eq:TPE}
\frac{\partial f}{\partial t} = - \nabla \cdot \left( \mu v \hat{b} f \right) &-& \frac{\partial}{\partial \mu} \left( \frac{1-\mu^2}{2L} vf \right) 
+ \frac{\partial}{\partial \mu} \left(D_{\mu\mu}  \frac{\partial f}{\partial \mu} \right)\nonumber \\  
&+&   \nabla \cdot \left( \mathbf{D}^{(x)}_{\perp}\cdot \nabla f \right) 
\end{eqnarray}

and solved by means of the numerical approach outlined by \citet{straussfichtner2015} to yield the gyro-tropic particle distribution function $f$. As we are simulating the propagation of electrons over very short time-scales {(a single event may last a few hours)}, we may safely neglect both adiabatic energy losses and co-rotation of the magnetic field. In Eq. \ref{Eq:TPE}, $\hat{b}$ is a unit vector pointing along the mean heliospheric magnetic field, {$v$ is the particle speed, $\mu$ is the cosine of the pitch-angle}, $ \mathbf{D}^{(x)}_{\perp}$ contains the perpendicular diffusion coefficients and is specified in spherical coordinates (radial distance, $r$, and azimuthal angle, $\phi$), $D_{\mu \mu}$ is the pitch-angle diffusion coefficient and the focusing length is calculated as 

\begin{equation}
L^{-1} = \nabla \cdot \hat{b}.
\end{equation}

Once Eq. \ref{Eq:TPE} is solved to obtain $f$, we also calculate the omni-directional intensity

\begin{equation}
F(r,\phi,t) =\frac{1}{2} \int_{-1}^{+1}   f(r,\phi,\mu,t)  d\mu
\end{equation}

and the first order anisotropy

\begin{equation}
A(r,\phi,t) = 3 \frac{\int_{-1}^{+1}  \mu f  d\mu}{\int_{-1}^{+1} f   d\mu},
\end{equation}

as these quantities can be compared directly to observations. 

As a boundary condition, the following isotropic injection function

\begin{equation}
f(r=r_0,\phi,t) =\frac{C}{t} \exp \left[ -\frac{\tau_a}{t}  -\frac{t}{\tau_e}  \right]
   \exp \left[ - \frac{(\phi - \phi_0)^2}{2 \sigma^2}  \right]  \label{Eq:reid_injection}
  \end{equation}

is prescribed at the inner boundary, located at $r_0 = 0.05$ AU. Gaussian injection in $\phi$ is assumed with $\phi_0 = \pi/2$ and $\sigma$ determining the broadness thereof. The value of $\sigma$ is varied in later sections. A Reid-Axford \citep[][]{reid1964} temporal profile is specified with $\tau_a = 1/10$ hr, $\tau_e = 1$ hr, and $C$ a constant. Other important quantities assumed in the model are: a \citet{parker} heliospheric magnetic field (HMF) normalized to 5 nT at Earth, a solar wind number density of 5 particles.cm$^{-3}$ at Earth decreasing as $r^{-2}$, and a constant solar wind speed of $V_{sw}=400$ km.s$^{-1}$. The pitch-angle and perpendicular diffusion coefficients, needed as input to Eq. \ref{Eq:TPE}, are discussed and calculated in the next section.

\section{Transport coefficients}

Charged particles are scattered by turbulent irregularities present in the solar wind, leading to diffusion both parallel and perpendicular to the mean field. To include the effect of turbulence, it is useful to decompose the resulting magnetic field, $\vec{B}$, into a locally uniform background field, $\vec{B_0} $, and a random turbulent component, $\vec{b}(x,y,z)$, such that

\begin{equation}
\vec{B} = \vec{B_0} +  \vec{b}(x,y,z),
\end{equation}

where $\langle \vec{B} \rangle= \vec{B_0}$ over long periods and $ \delta B^2=\langle \vec{b}^{2} \rangle$ the variance thereof. To correctly characterize the solar wind turbulence, and couple this to particle scattering, remains one of the biggest challenges in heliospheric physics.

%

Following \citet{shalchibook}, we assume the fluctuating field to be separable into a {\it slab} and {\it 2D} component \citep[to be defined later; see also][]{Matthaeusetal1995}, so that the turbulence power spectrum in wavenumber-space for homogeneous composite turbulence is,

\begin{equation}
\label{Eq:shalchi_pluh_pluh}
\mathcal{P} (\vec{k},t) = g^{\mathrm{slab}} (k_{||}) \Gamma^{\mathrm{slab}} (k_{||},t) + g^{\mathrm{2D}} (k_{\perp}) \Gamma^{\mathrm{2D}} (k_{\perp},t),
\end{equation}

and $\Gamma$ being the dynamical correlation function, which may be different for each turbulent component. For the total variance of the fluctuations, it follows that

\begin{equation}
\delta B^2 = \delta B_{\mathrm{slab}}^2 + \delta B_{\mathrm{2D}}^2.
\end{equation}

In order to calculate the transport coefficients, we make use of the so-called {\it Shalchi slab hypothesis}, where it is assumed that only slab turbulence will influence parallel particle transport, while the 2D component will lead, exclusively, to perpendicular diffusion \citep[][]{Shalchi2006}. This is motivated by the simulations of \citet{qin2002}, showing that the slab component does not contribute significantly to perpendicular diffusion, while the contribution to pitch-angle scattering from perpendicular waves (which can be modeled as part of the 2D component) is usually considered to be small \citep[e.g.][]{schlikkie}.

\begin{deluxetable*}{lll}
\tabletypesize{\footnotesize}
\tablecolumns{7} 
\tablewidth{0pt} 
\tablecaption{The turbulence quantities employed in this study. See text for a description of the various quantities listed below.}
\tablehead{\colhead{Turbulence quantity}                                          & 
           \colhead{Value or expression adopted}                                          & 
           \colhead{Reference}                                          }
\startdata
$\delta B^2 (r=1 \mathrm{AU})$  & $13.2$ nT$^2$ & \citet{bieberetal1994}         \\
$\delta B^2 $  & $\sim r^{-2.4}$ & \citet{EB2013}       \\
$s$  & $5/3$ &   Kolmogorov decay      \\
$p$  & $2.6$ & \citet{Smith2006b}        \\
$k_{min}$  & $35$ AU$^{-1}$ & \citet{Weygand2011}  \tablenotemark{a}     \\
$k_d$  & $2 \pi \left(a + b \Omega_i  \right)/V_{sw}$ & \citet{leamonetal2000} \tablenotemark{b}      \\
$\delta B^2_{\mathrm{slab}}$  & $0.2 \delta B^2$  &\citet{bieberetal1994}        \\
$q$  & $7$ &    \citet{2007_Mattheaus_etal_ApJ} \tablenotemark{c}     \\
$\nu$  & 5/3 &    Kolmogorov decay      \\
$k_{2D}$  & $135$ AU$^{-1}$ & \citet{Weygand2011}  \tablenotemark{d}     \\
$k_{out}$  & $k_{2D}/100$   &  \citet{EB2015} \tablenotemark{e}      \\
$\delta B^2_{\mathrm{2D}}$  & $0.8 \delta B^2$  & \citet{bieberetal1994}       \\
 \\
\enddata

\tablenotetext{a}{The value of $k_{min}$ was chosen such that the resulting slab correlation length compares to the values reported by \citet{Weygand2011}.}

\tablenotetext{b}{Assuming $k_d \propto \Omega_i$, where $\Omega_i$ is the proton gyrofrequency, and using the best fit regression values of $a=0.2$ Hz and $b=1.76$.}

\tablenotetext{c}{\citet{2007_Mattheaus_etal_ApJ} show that $q$ should be an odd integer and that $q > 1$.}

\tablenotetext{d}{We assume, for the 2D correlation length $\langle l_{\perp} \rangle$, that $\langle l_{\perp} \rangle = 1 /k_{2D}$, and use the observations of \citet{Weygand2011} to constrain $\langle l_{\perp} \rangle$. }

\tablenotetext{e}{\citet{EB2015} considers this value a reasonable estimate as no {\it in-situ} observations of the outer scale exist at present.}

\end{deluxetable*}

\begin{figure*}
\includegraphics[width=160mm]{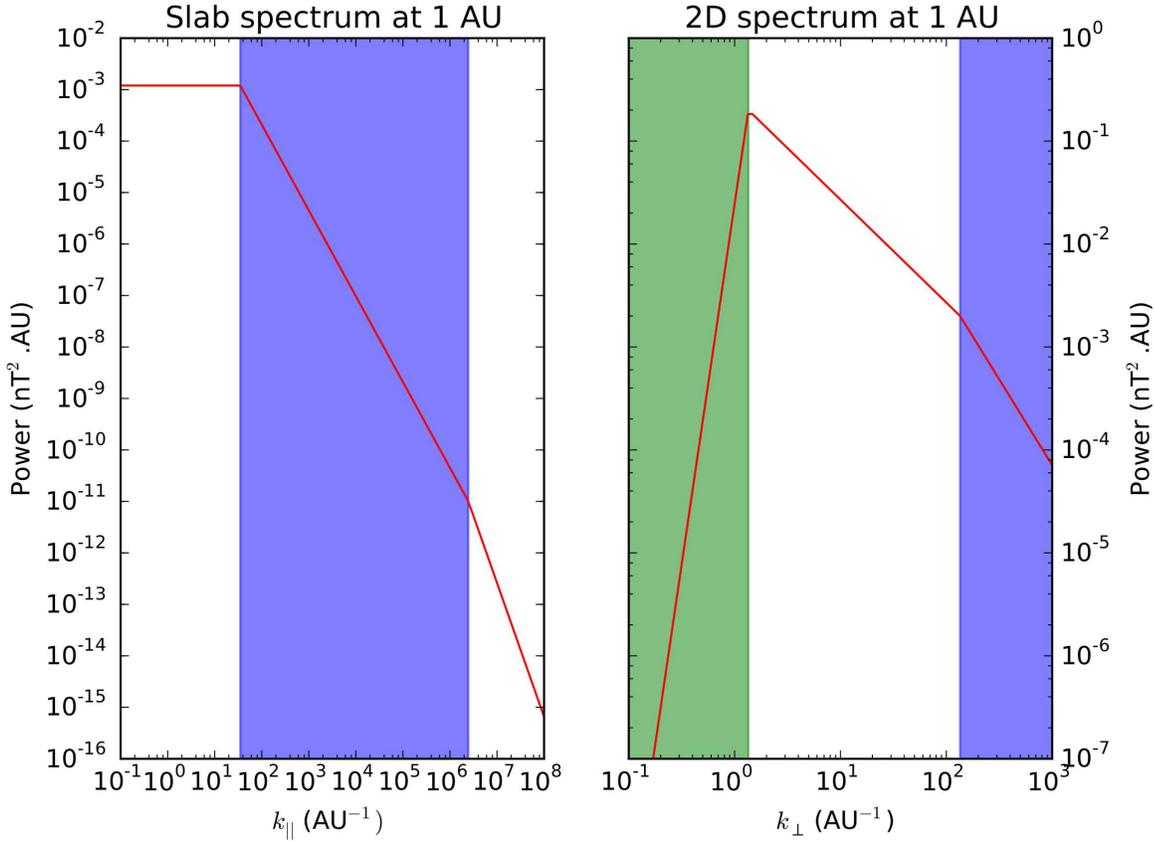}
\caption{The turbulence spectra used in this study. The blue shaded regions show the inertial ranges, while the green region indicates the ``outer range" of the 2D spectrum. \label{fig1}}
\end{figure*}

\subsection{Slab turbulence and pitch-angle diffusion}

\begin{figure*}
\includegraphics[width=160mm]{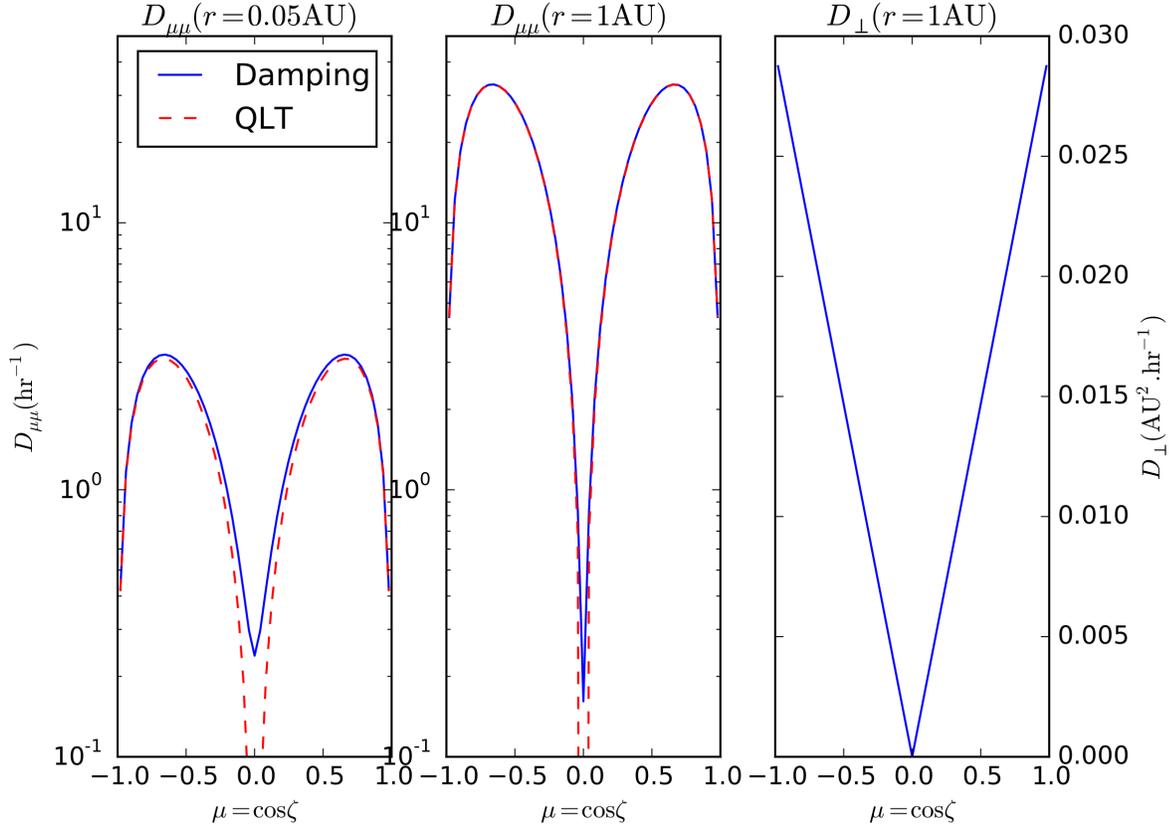}
\caption{The {calculated} $D_{\mu \mu}$ at 0.05 AU (left panel) and 1 AU (Earth; middle panel) and $D_{\perp}$ at 1 AU (Earth; right panel) as a function of $\mu$. For $D_{\mu \mu}$, two scenarios are shown at each radial position, namely, using the damping function (solid blue lines) and using standard QLT (dashed red lines). \label{fig2}}
\end{figure*}

For the slab component (fluctuations with wave-vectors parallel to the mean field, i.e. $k_{||}$), we use the plasma wave model of slab turbulence \citep[][]{schlikkie}, where

\begin{equation}
\Gamma (k_{||},t) = \exp \left( i \omega t - \gamma t \right).
\end{equation}

The wave damping rate is given by $\gamma$ and $\omega$ is the wave frequency, coupled to $k_{||}$ via the wave dispersion relation, which, for non-dispersive Alfv{\'e}n waves is

\begin{equation}
\omega = j V_A k_{||},
\end{equation}

where $j =\pm$ labels forward $(j=+1)$ or backward $(j=-1)$ propagating waves {and $V_A$ is the Alfv\'en speed}. The strength of the slab component is calculated as

\begin{equation}
\label{Eq:slab_normalize}
\delta B_{\mathrm{slab}}^2 =  8 \pi \int_{0}^{\infty}  g^{\mathrm{slab}}(k_{||}) dk_{||},
\end{equation}

with $g^{\mathrm{slab}}(k_{||})$ the omni-directional one-sided slab spectrum. In deriving Eq. \ref{Eq:slab_normalize}, the slab component was assumed to be axisymmetric, and hence, a zero cross helicity is assumed throughout. Following \citet{ts2003}, we choose the spectral form of the slab spectrum as

\[ g^{\mathrm{slab}}(k_{||})=g_{0}^{\mathrm{slab}}  \left\{
\begin{array}{ll}
      k_{\mathrm{min}}^{-s} & 0 \leq k_{||} \leq k_{\mathrm{min}} \\
      k_{||}^{-s} & k_{\mathrm{min}} < k_{||} < k_{\mathrm{d}} \\
      k_{\mathrm{d}}^{p-s} k_{||}^{-p} & k_{||} \geq k_{\mathrm{d}} \\
\end{array} 
\right. \]

which contains an energy range which is independent of $k_{||}$ below $k_{\min}$, an inertial range between $k_{\min}$ and $k_d$ and a dissipation range above $k_d$, and $g_{0}^{\mathrm{slab}}$ determined from the normalization condition, Eq. \ref{Eq:slab_normalize}, to give 

\begin{equation}
g_0^{\mathrm{slab}} = \frac{\delta B^2_{\mathrm{slab}}}{8\pi} (s - 1) k_{\mathrm{min}}^{s - 1} \left[ s  + \frac{s - p}{p - 1} \left( \frac{k_{\mathrm{min}}}{k_{\mathrm{d}}}  \right)^{s - 1} \right] ^{-1}.
\end{equation}

The turbulence quantities used in this study are listed in Table 1, while the resulting turbulence spectra, at 1 AU, are shown in Fig. \ref{fig1}.

The resulting plasma wave pitch-angle diffusion coefficient, for vanishing cross-helicity and magnetic helicity is given by \citep[][]{schlikkie}

\begin{eqnarray}
\label{Eq:dmumu_general}
D_{\mu \mu} &=& \frac{2 \pi v^2 \left( 1 - \mu^2 \right) }{B_0^2 r_L^2} \left[ 1 - \frac{ \mu \omega}{k_{||} v} \right]^2  
\int_0^{\infty} g^{\mathrm{slab}} (k_{||}) \nonumber \\ 
& \times &  \left[ \mathcal{R}_{n=+1} (k_{||}) + \mathcal{R}_{n=-1} (k_{||})  \right] d k_{||}
\end{eqnarray}

where $\mathcal{R}_{n= \pm 1}$ are resonance functions related to the two possible polarization states of the wave turbulence component,

\begin{equation}
\label{Eq:general_resonances}
\mathcal{R}_{n=\pm 1}(k_{||}) = \frac{\gamma}{\gamma + \left( v\mu k_{||} - \omega \pm \Omega \right)^2}, 
\end{equation}

and where we have assumed the damping rate to be the same for both wave polarizations of Alfv{\'e}n waves. For $v \gg V_A$, we may drop the term 

\begin{equation}
\left[ 1 - \frac{ \mu \omega}{k_{||} v} \right]^2 = \left[ 1 - \mu \frac{  V_A}{ v} \right]^2 \approx 1.
\end{equation}

We apply the fast-particle assumption in the form $\omega \ll \Omega$, with $\Omega$ the particle gyro-frequency, so that, 

\begin{equation}
\label{Eq:fast_particle_resonances}
\mathcal{R}_{n=\pm 1}(k_{||}) \approx \frac{\gamma}{\gamma + \left( v\mu k_{||}  \pm \Omega \right)^2}.
\end{equation}

If damping is neglected, $\gamma \rightarrow 0$, Eq. \ref{Eq:fast_particle_resonances} reduces to

\begin{equation}
\mathcal{R}_{n=\pm 1} \rightarrow \pi \delta \left( v\mu k_{||}  \pm \Omega \right),
\end{equation}

which, together with Eq. \ref{Eq:dmumu_general}, leads to the well known quasi-linear theory (QLT) result for $D_{\mu \mu}$ \citep[][]{jokipii1966,qinshalchi2009} ,

\begin{equation}
\label{Eq:dmumu_QLT}
D_{\mu \mu}^{\mathrm{QLT}} = \frac{2 \pi^2 v \left( 1 - \mu^2 \right) }{ |\mu| B_0^2 r_L^2}    g^{\mathrm{slab}} \left( k_{||}^{\mathrm{res}}  \right) ,
\end{equation}

with the resonant wavenumber 

\begin{equation}
k_{||}^{\mathrm{res}} = \frac{1}{|\mu| r_L}.
\end{equation}

We will however evaluate the {\it dynamical} (or, damped) turbulence scenario where $\gamma \neq 0$, and use

\begin{eqnarray}
\label{Eq:dmumu_we_use}
D_{\mu \mu}^{\mathrm{DT}} &=& \frac{2 \pi v^2 \left( 1 - \mu^2 \right) }{B_0^2 r_L^2}   \int_0^{\infty} g^{\mathrm{slab}} (k_{||})  \\
 &\times&\left[ \frac{\gamma}{\gamma + \left( v\mu k_{||}  + \Omega \right)^2} + \frac{\gamma}{\gamma + \left( v\mu k_{||}  - \Omega \right)^2}  \right] d k_{||} \nonumber,
\end{eqnarray}

where, for ease of calculation, the integral is evaluated numerically. Following \citet{bieberetal1994}, we use a damping rate with the form of

\begin{equation}
\gamma = \alpha \omega = \alpha V_A k_{||},
\end{equation}

where $\alpha \in [0,1]$ is a constant determining the level of damping. Unless otherwise specified, we use $\alpha = 1$ for maximum effects. The pitch-angle dependence of $D_{\mu \mu}$ is shown in Fig. \ref{fig2} at $r=0.05$ AU (left panel) and at $r=1$ AU (middle panel) for the scenario with damping (solid lines) and standard QLT (dashed lines). The effect of resonance broadening via a damping process is clear: a finite value of $D_{\mu \mu}$ at $\mu \sim 0$ is obtained as required by observations. Moreover, because $V_A$ becomes increasingly larger near the Sun, the damping rate increases, leading to increased resonance broadening in these regions (as compared to QLT).

The parallel mean free path (MFP, $\lambda_{||}$) is calculated from $D_{\mu \mu}$ following the usual definition of \citet{hasselmaanwibberenz1968},

\begin{equation}
\lambda_{||} = \frac{3v}{8} \int_{-1}^{+1} \frac{\left( 1 - \mu^2 \right)^2}{D_{\mu\mu}} d\mu .
\end{equation}

The resulting $\lambda_{||}$ is shown in Fig. \ref{fig3} as a function of radial distance. Also shown on the figure are ``consensus" values of $\lambda_{||}$ at Earth, ranging from $\lambda_{||}=0.24$ \citep[][]{wolfgang,drogeteal2014} to $\lambda_{||}=1$ AU \citep[][and references therein]{bieberetal1994} {for electrons with an energy of $\sim 100$ keV}. Newer estimates by \citet{drogeteal2016} indicate $\lambda_{||} \sim 0.15 - 0.6$ AU. Our set of assumed parameters result in $\lambda_{||} \sim 0.7$ AU, therefore falling nicely into this range of accepted values. Also included in Fig. \ref{fig3} is the focusing length ($L$, green line). Generally, in regions where $\lambda_{||} \gg L$, anisotropic behavior is expected, and hence, our simulations near Earth are expected to give an almost beam-like (i.e. highly anisotropic) distribution of electrons. The radial dependence of $\lambda_{||}$ does appear more complex than generally thought: Previous studies have assumed that $\lambda_{||}$ should increase linearly with radial distance. Although our calculated $\lambda_{||}$ does exactly this beyond 1 AU, we note a large increase of $\lambda_{||}$ near the Sun. This complex radial dependence is also seen in the estimates of \citet{Laitinenetal2016}.
\subsection{2D turbulence and perpendicular diffusion}

\begin{figure}
\includegraphics[width=80mm]{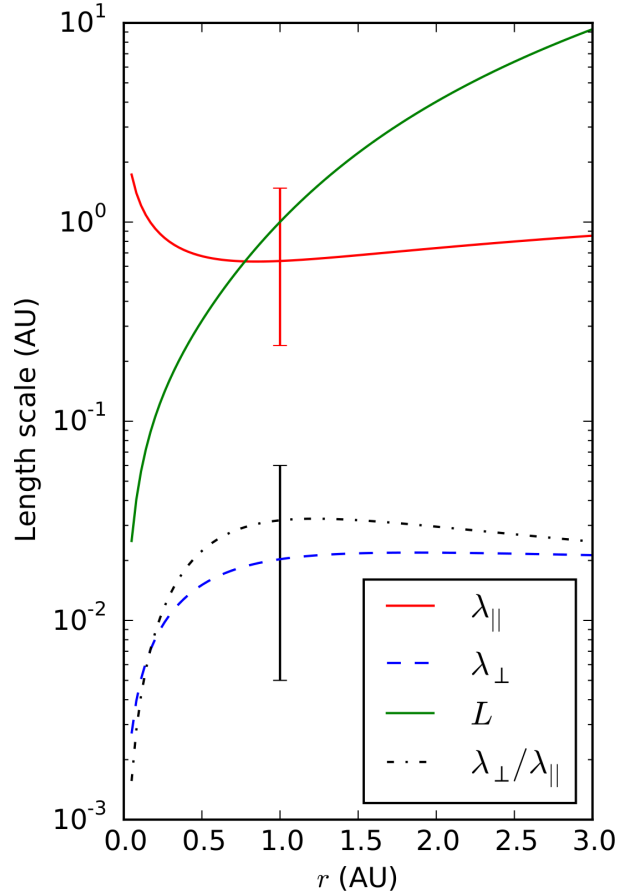}
\caption{The parallel (red line) and perpendicular (dashed blue line) mean free path as a function of radial distance. Also shown is the focusing length (solid green line) and the ratio $\eta$ (note that this quantity does not have any units) as the dash-dotted black line. The vertical red and black error intervals show estimates for $\lambda_{||}$ and $\eta$ respectively. For the calculation of $\lambda_{\perp}$, we used $a=1/10$. \label{fig3}}
\end{figure}

We assume a 2D component (that is, fluctuations with wave-vectors perpendicular to the mean field, $k_{\perp}$) that is magnetostatic (in Eq. \ref{Eq:shalchi_pluh_pluh} this implies $\Gamma^{2D} (k_{\perp}, t) = 1$) with the strength of the 2D component calculated as

\begin{equation}
\label{Eq:2D_normalize}
\delta B^2_{\mathrm{2D}} =  2\pi \int_{0}^{\infty} g^{\mathrm{2D}}(k_{\perp}) dk_{\perp}.
\end{equation}

For the 2D spectrum we choose a form, similar to that employed {by} \citet{EB2015}, given by

\[ g^{\mathrm{2D}}(k_{\perp})=g_{0}^{\mathrm{2D}}  \left\{
\begin{array}{ll}
      \frac{k_{\mathrm{2D}}}{k_{\mathrm{out}}} \left( \frac{k_{\perp}}{k_{\mathrm{out}}} \right)^{q} & 0 \leq k_{\perp} \leq k_{\mathrm{out}} \\
      \left( \frac{k_{\perp}}{k_{\mathrm{2D}}} \right)^{-1} & k_{\mathrm{out}} < k_{\perp} < k_{\mathrm{2D}} \\
      \left( \frac{k_{\perp}}{k_{\mathrm{2D}}} \right)^{-\nu} & k_{\perp} \geq k_{\mathrm{2D}} \\
\end{array} 
\right. \]

with a so-called inner-range below $k_{out}$ as required by \citet{2007_Mattheaus_etal_ApJ}, an energy range that decreases as $k_{\perp}^{-1}$ and an inertial range beyond $k_{2D}$. Eq. \ref{Eq:2D_normalize} is again used to determine $g_{0}^{\mathrm{2D}}$ and leads to

\begin{equation}
g_0^{\mathrm{2D}} = \frac{\delta B^2_{\mathrm{2D}}}{2\pi k_{\mathrm{2D}}} \left[ \frac{1}{q+1} + \ln \left( \frac{k_{\mathrm{2D}}}{k_{\mathrm{out}}} \right)  + \frac{1}{\nu -1}  \right]^{-1}.
\end{equation}

The turbulence quantities used in this study are summarized in Table 1, while the resulting turbulence spectra at 1 AU are shown in Fig. \ref{fig1}.

For $D_{\perp}$, we implement the field-line random walk (FLRW) model of \citet{jokipii1966}, where

\begin{equation}
D_{\perp} = a v |\mu| \kappa_{\mathrm{FL}},
\end{equation}

and the field-line diffusion coefficient $\kappa_{\mathrm{FL}}$ is given by e.g. \citet{qinshalchi2014} as

\begin{equation}
\kappa^2_{\mathrm{FL}} = \frac{\pi}{B_0^2} \int_0^{\infty} k^{-2}_{\perp} g^{\mathrm{2D}} (k_{\perp})  dk_{\perp}.
\end{equation}

\citet{qinshalchi2014} note that the above result is equivalent to the FLRW diffusion coefficient in the presence of pure 2D axisymmetric turbulence. This is given by \citet{2007_Mattheaus_etal_ApJ} as

\begin{equation}
\kappa_{\mathrm{FL}} = \frac{\sqrt{\delta B_{2D}^{2}/2}}{B_{o}}\lambda_{u}
\end{equation}

where $\lambda_{u}$ denotes the 2D ultrascale, which can be calculated for the spectral form used in this study using \citep{2007_Mattheaus_etal_ApJ}

\begin{equation}
\lambda_{u} = \sqrt{\frac{\int k^{-2}S(k)d^{2}k}{\delta B_{2D}^{2}}} = \sqrt{\frac{2\pi\int k^{-2}_{\perp} g^{\mathrm{2D}} (k_{\perp})  dk_{\perp}}{\delta B_{2D}^{2}}}
\end{equation}

with $S(k)$ the 2D modal spectrum \citep[see, e.g.,][]{batchelor,2007_Mattheaus_etal_ApJ}. This then yields

\begin{equation}
\lambda_{u} = \sqrt{\frac{\left[ k^{-2}_{\mathrm{out}}\left( \frac{q+1}{2q - 2} \right)  + k^{-2}_{\mathrm{2D}}\left( \frac{1 - \nu}{2\nu + 2}  \right) \right]}{\left[ \frac{1}{q+1} + \ln \left( \frac{k_{\mathrm{2D}}}{k_{\mathrm{out}}} \right)  + \frac{1}{\nu -1}  \right]}}.
\end{equation}

Either way, the resulting FLRW perpendicular diffusion coefficient is then

\begin{equation}
\kappa^2_{\mathrm{FL}} = \frac{1}{2} \frac{\delta B_{\mathrm{2D}}^2}{B_0^2} \frac{\left[ \frac{1}{k^2_{\mathrm{out}}}                                  
 \left( \frac{q+1}{2q - 2} \right)  +\frac{1}{k^2_{\mathrm{2D}}} \left( \frac{1 - \nu}{2\nu + 2}  \right)      \right]}{\left[ \frac{1}{q+1} + \ln \left( \frac{k_{\mathrm{2D}}}{k_{\mathrm{out}}} \right)  + \frac{1}{\nu -1}  \right]}.
\end{equation}

From $D_{\perp}$ we can calculate $\lambda_{\perp}$, the {isotropic perpendicular mean free path}, as

\begin{equation}
\lambda_{\perp} = \frac{3}{2v} \int_{-1}^{+1} D_{\perp} d \mu.
\end{equation}

{The parameter $a$ was originally introduced into the non-linear guiding center theory (NLGC) as part of an Ansatz \citep[see][]{shalchibook} to describe the components of the velocity of the particle gyrocenter in terms of the velocity $v_z$ of the gyrocenter parallel to the uniform background magnetic field $B_0$, and the relevant fluctuating component of the magnetic field, such that, for example,

\begin{equation}
v_x = a v_{z}\frac{b_{x}}{B_0  } \label{schmeh}.
\end{equation}

The exact interpretation of $a$, as well as its value, has been somewhat unclear in the past, it being generally interpreted as being related to the probability that particles follow field lines \cite[see, e.g.][]{shaolchi2010,hs2014}. Theoretical calculations, most notably a derivation of NLGC from the Newton-Lorentz equation by \citet{sd2008}, indicated that $1 \leq a^2 \leq 2$, whereas numerical test-particle simulations such as those performed by \citet{Matthaeusetal2003} required a value smaller than unity. \citet{ shalchi2016}, however, by investigating the effects of the assumption of a finite gyroradius on the analytical theory of perpendicular diffusion, find that $a^2$ can be interpreted as a parameter describing the finite gyroradius effects of turbulence on the particle. Furthermore, \citet{ shalchi2016} finds that these effects lead to $a^{2} \leq 1$. In this work, we treat $a \in [ 0,1  ]$ as a free parameter and illustrate the effect of this changing value on modeled intensities of SEPs in later sections.}

The form of $D_{\perp}$, although important for modeling studies \citep[e.g.][]{straussfichtner2014,straussfichtner2015} is, however, not well known, as most studies have focused on deriving the isotropic perpendicular scattering coefficient \citep[e.g.][]{Matthaeusetal2003}. Without a clear second-order, non-linear framework for $D_{\perp}$, we opt to use the FLRW approach in this work. For low energy electrons, however, this is most likely not a bad approximation: at Earth, for 100 keV electrons, we have $r_L \sim 10^{-7} $ AU, and $\langle l_{\perp} \rangle  \sim 0.001$ AU for the perpendicular correlation length, so that we may comfortably work in the limit of $r_L \ll \langle l_{\perp} \rangle$, for which the FLRW coefficient is a good approximation \citep[][]{straussetal2016}. A derivation of $D_{\perp}$, on the pitch-angle level, is an ongoing topic of investigation \citep[e.g.][]{frashettijokipii,qinshalchi2014,frashetti}. When $r_L$ becomes comparable with or larger than $ \langle l_{\perp} \rangle$, a non-linear theory is needed to correctly describe perpendicular diffusion. \citet{drogeteal2016}, using a phenomenological approach, recently found $\lambda_{\perp}$ to be independent of $\lambda_{||}$ (see their Fig. 17), which is an additional indication that the FLRW coefficient is a good approximation for low energy electrons.

The pitch-angle dependence of $D_{\perp}$ is shown in the right panel of Fig. \ref{fig2}, while the radial dependence of $\lambda_{\perp}$ is shown in Fig. \ref{fig3} using $a=1/10$. The effectiveness of perpendicular diffusion is usually quantified by the ratio $\eta = \lambda_{\perp}/\lambda_{||}$, which is generally believed to be in the range of $\eta \in (0.005,0.05)$ \citep[see, amongst others, the discussion by][]{shalchibook}. This ratio, together with its consensus values, is also indicated on Fig. \ref{fig3}. Similarly to $\lambda_{||}$, the radial dependence of $\lambda_{\perp}$ is much more complex than originally thought, decreasing significantly towards the Sun. Beyond 1 AU, $\lambda_{\perp}$ assumes approximately constant levels. The ratio $\eta$ is therefore also not a constant as assumed in most modeling studies. Again, this complicated radial dependence is similar to that implemented by \citet{Laitinenetal2016}.

\section{Results}

As described in the proceeding sections, our model set-up contains two free parameters, namely the broadness of the injected boundary condition (given by $\sigma$) and the effectiveness of perpendicular diffusion (determined by $a$). In this section, we show modeled intensities of SEPs for various combinations of these parameters.

\subsection{General results}

\begin{figure*}
\includegraphics[width=180mm]{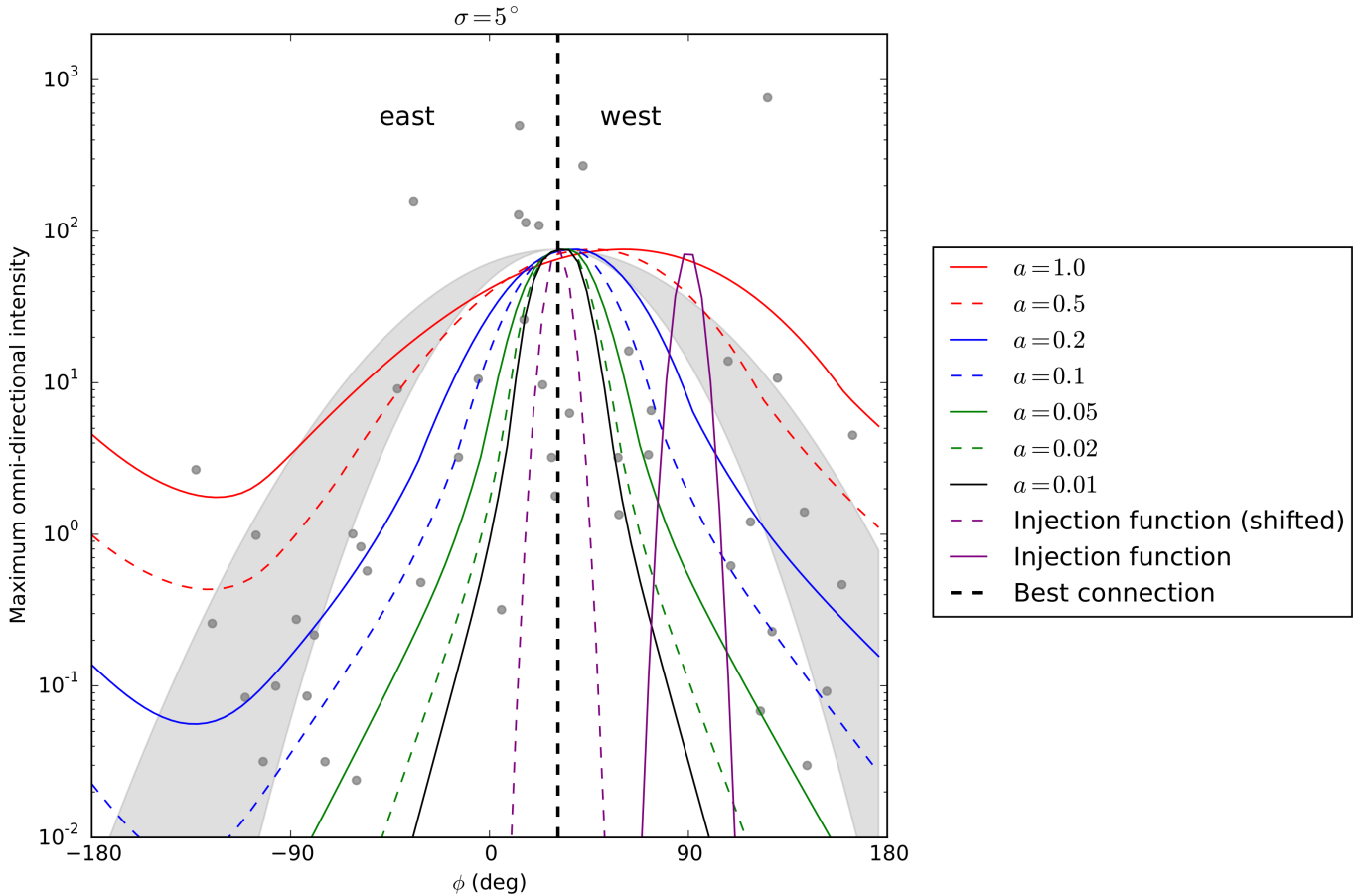}
\caption{The modeled maximum omni-directional intensity as a function of longitude for different levels of perpendicular diffusion (indicated by $a$). The solid purple line shows the injection function, with $\sigma = 5^{\circ}$, specified at the inner boundary, while the dashed purple line shows this distribution shifted to the position of best magnetic connection at Earth ($\phi \approx 30^{\circ}$, indicated by the vertical dashed line). {The grey symbols and band are observed electron peak intensities in the range of 55 -- 105 keV and the range of corresponding Gaussian fits of these multi-spacecraft events taken from \citet{dresing2014}}. \label{fig4}}
\end{figure*}

\begin{figure*}
\includegraphics[width=160mm]{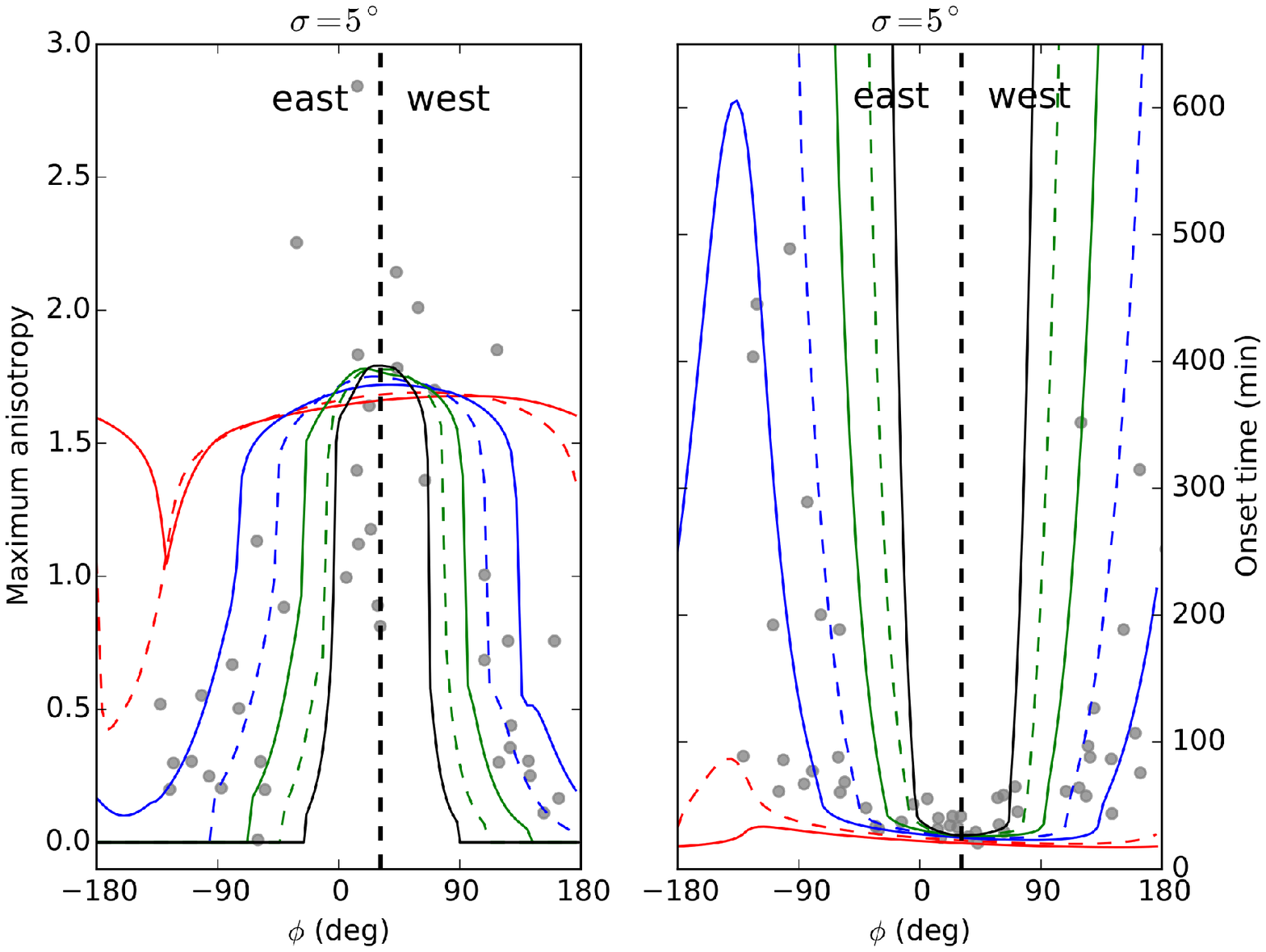}
\caption{The computed maximum anisotropy (left panel) and onset time (right panel) for different levels of perpendicular diffusion (see again the legend of Fig. \ref{fig4} where the lines correspond to different values of $a$). The observations from \citet{dresing2014} are shown as grey symbols. \label{fig5}}
\end{figure*}

We start by modeling the peak (maximum) omni-directional intensity as a function of longitude at Earth's orbit ($r = 1$ AU). The results, for a constant injection broadness of $\sigma = 5^{\circ}$ and varying levels of perpendicular diffusion, are shown in Fig. \ref{fig4}. {The value of $\sigma = 5^{\circ}$ is chosen purely for demonstration purposes and can, of course, be changed in future.} On this figure, the {injected distribution (normalized to the resulting maximum differential intensity at Earth)} is shown as the narrow Gaussian distribution peaking at $\phi = 90^{\circ}$ (the solid purple line on the figure) at the inner model boundary. If perpendicular diffusion was to be completely absent, this distribution would simply be shifted in longitude, as it follows the curved interplanetary field, and be observed at Earth at the dashed purple distribution, peaking at $\phi \approx 30^{\circ}$. In our model set-up this corresponds to the longitude, at Earth's position, that is magnetically connected to the peak of the injection function at the inner boundary. This so-called ``longitude of optimal magnetic connection" is indicated on this, and later figures, by a vertical dashed line. Following our usual definition $\phi$ (see \citet{straussfichtner2015}, especially their Fig. 3), $\phi \gtrsim 30 ^{\circ}$ corresponds to western longitudes and $\phi \lesssim 30 ^{\circ}$ to eastern longitudes (as also indicated on the figure).

We can now increase the effectiveness of perpendicular diffusion in the model by increasing the value of $a$, up to its maximum value of $a=1$. As shown in Fig. \ref{fig4}, this results, as expected, in broader particle distributions with the broadness increasing with the value of $a$. The modeled intensities are compared to observations reported by \citet{dresing2014}, these being shown as the grey symbols while their Gaussian fit, with a broadness of $\sigma = 35^{\circ} \rightarrow 48^{\circ}$, is shown as the grey filled band. {Note that \citet{dresing2014} only selected so-called widespread events in their study.} A similar study by \citet{larioetal2013}{, which studied multi-spacecraft events regardless of their longitudinal extent,} estimates that $\sigma \approx 49^{\circ}$, which is consistent with the upper limits of the \citet{dresing2014} observations if we ignore any possible asymmetries in the distribution (see Sec. \ref{Sec:assymmetries} for a detailed discussion regarding this topic). {Although we do not attempt a detailed reproduction of the observations, which show large inter-event variability, Fig. \ref{fig4} does illustrate the ability of the model to reproduce the observed broadness of the SEP distribution through fundamentally prescribed diffusion processes.} Moreover, we note that a value of $a=0.2-0.5$ seems to be consistent with the observed band, while Fig. \ref{fig3} reveals that these values of $a$ result in $\lambda_{\perp}$'s that are still well within the consensus range thereof.

A second set of observable quantities are calculated and shown in Fig. \ref{fig5} as a function of longitude at Earth's position. They are the maximum anisotropy (left panel) and the onset time ({also referred to as the onset delay,} right panel). Again we assume $\sigma = 5^{\circ}$ and vary the value of $a$ (the different solutions correspond to the legend of Fig. \ref{fig4}). In order to calculate both of these quantities, we need to specify an artificial background intensity in the model, mimicking a SEP event breaching such a background level. For this study, we set the background level as $1/1000$-th of the maximum intensity. A similar approach was used {by} \citet{wangqin} and \citet{straussfichtner2015}. The onset time is calculated as the time it took the modeled intensities to breach this background level, while the maximum anisotropy is only registered for times longer than the onset time in order to be comparable with observations. The fact that the onset times may depend on the assumed background is definitely not optimal, but without a better way to quantify the temporal shape of a SEP event, it remains the preferred method to do so (this is, of course, also true in the experimental case, see e.g. \citet{xieetal2016}). In Fig. \ref{fig5}, we compare our results to the observations of \citet{dresing2014}, {with only the anisotropy values of the so-called ``class 1" and ``class 2" events shown; these are widespread events where either perpendicular diffusion or an extended source are believed to play a large role.}

From Fig. \ref{fig5}, we note that both of the calculated quantities are extremely sensitive to the level of perpendicular diffusion, with the shortest onset times and largest anisotropies generally seen at optimal magnetic connection to the source, the idea being that these particles are basically ``free-streaming" from the source to the observer. Away from these longitudes, perpendicular diffusion plays a more evident role in  spreading the SEPs in longitude, but, of course, taking a finite amount of time to do so, resulting in much longer onset times. The longer onset times, due to longer propagation times, generally result in much more isotropic distributions away from best magnetic connection. In the case of extremely effective perpendicular diffusion ($a=1$), it is however possible to get large anisotropies over almost all longitudes due to the fact that the FLRW coefficient, which scales as $D_{\perp} \sim |\mu|$, transports a beam-like distribution (a highly anisotropic distribution with $\mu \sim 1$) most effectively in longitude. {The $a=1$ case therefore results in a SEP transport process that is entirely dominated by perpendicular diffusion so that SEPs with $\mu = 1$ are transported effectively in longitude (i.e. perpendicular to the mean field) without being scattered significantly in pitch-angle space.} This extreme case is however {unrealistic: this is also evident from e.g. the observed values of the onset time which show a much larger $\phi$ dependence than the modeling results with $a=1$, while the $\lambda_{\perp}$ calculated for such a scenario will be much larger than the consensus range (see again Fig. \ref{fig3}).} We note that, for both quantities shown in Fig. \ref{fig5}, a good fit is once again obtained when $a=0.2$ (the solid blue curve in both panels).

\subsection{Azimuthal asymmetries}
\label{Sec:assymmetries}

Much modeling and experimental work have recently touched on the symmetry of the SEP distribution at Earth. Naively, one would expect the distribution to approximate a Gaussian form (in terms of longitude), peaking at the longitude of best magnetic connection. Observationally, this has however not been {the case: \citet{larioetal2013} have found that, for electrons, the maximum is displaced, or shifted, by $\sim 16^{\circ}$ towards western longitudes}. A similar shift was reported for proton events by \citet{larioetal2006} and \citet{hewang2016}. In terms of modeling, such a shift can be explained in terms of a combination of co-rotation and perpendicular diffusion \citep[][]{giacalonejokipii} or by only implementing perpendicular diffusion \citep[][]{straussfichtner2015}. When examining the results presented in Fig. \ref{fig4}, we note a similar shift of the peak intensity towards western longitudes, with the shift increasing with increasing levels of perpendicular diffusion: for $a=0.2$, the shift is $\sim 10^{\circ}$, increasing to $\sim 40^{\circ}$ for $a=1.0$. As discussed in detail by \citet{straussfichtner2015}, such a shift in peak intensity can be explained in terms of perpendicular diffusion along a curved magnetic field, where, close to Earth, the ``perpendicular" direction also points radially away from the Sun (but only when magnetically connected towards the west of best connection to the source). This shift, albeit small, must therefore be a characteristic of all diffusion models. Indeed, subsequent models by \citet{hewang2015} and \citet{Ablassmayeretal} have also confirmed this idea. \footnote{\citet{kahler2016} mistakenly states that the results of \citet{straussfichtner2015} and \citet{hewang2015} are not consistent and that the shifts are directed towards different longitudes. We, however, confirm the consistency of previous model results with the peak intensity shifting to western longitudes. The confusion is probably related to the different definitions of longitude used in the different models.} 

As already mentioned, additional processes may also contribute to the asymmetrical nature of the SEP distribution, such as co-rotation in long-lasting events \citep[][]{larioetal2014} or drift effects \citep[][]{marshetal2013}, although the latter process would result in electron and proton events shifting towards different longitudes: an effect not yet observed. It is also questionable whether these small shifts in the peak intensity are actually observable given that the error in determining the foot-point of magnetic field lines, using the simple Parker model, is probably $\sim 10^{\circ}$ \citep[][]{kahleretsal2016}, and could, at times, be even worse \citep[][]{lietal2016}.

Finally, we note from the right panel of Fig. \ref{fig5} that the modeled onset times are also not symmetric and somewhat shorter for western longitudes, while the corresponding anisotropy is marginally {larger}. This can again be explained in terms of perpendicular diffusion, while observations from \citet{richardsonetal2014} hint at such an asymmetry being present in the observations.

\subsection{Temporal evolution}

\begin{figure*}
\includegraphics[height=60mm]{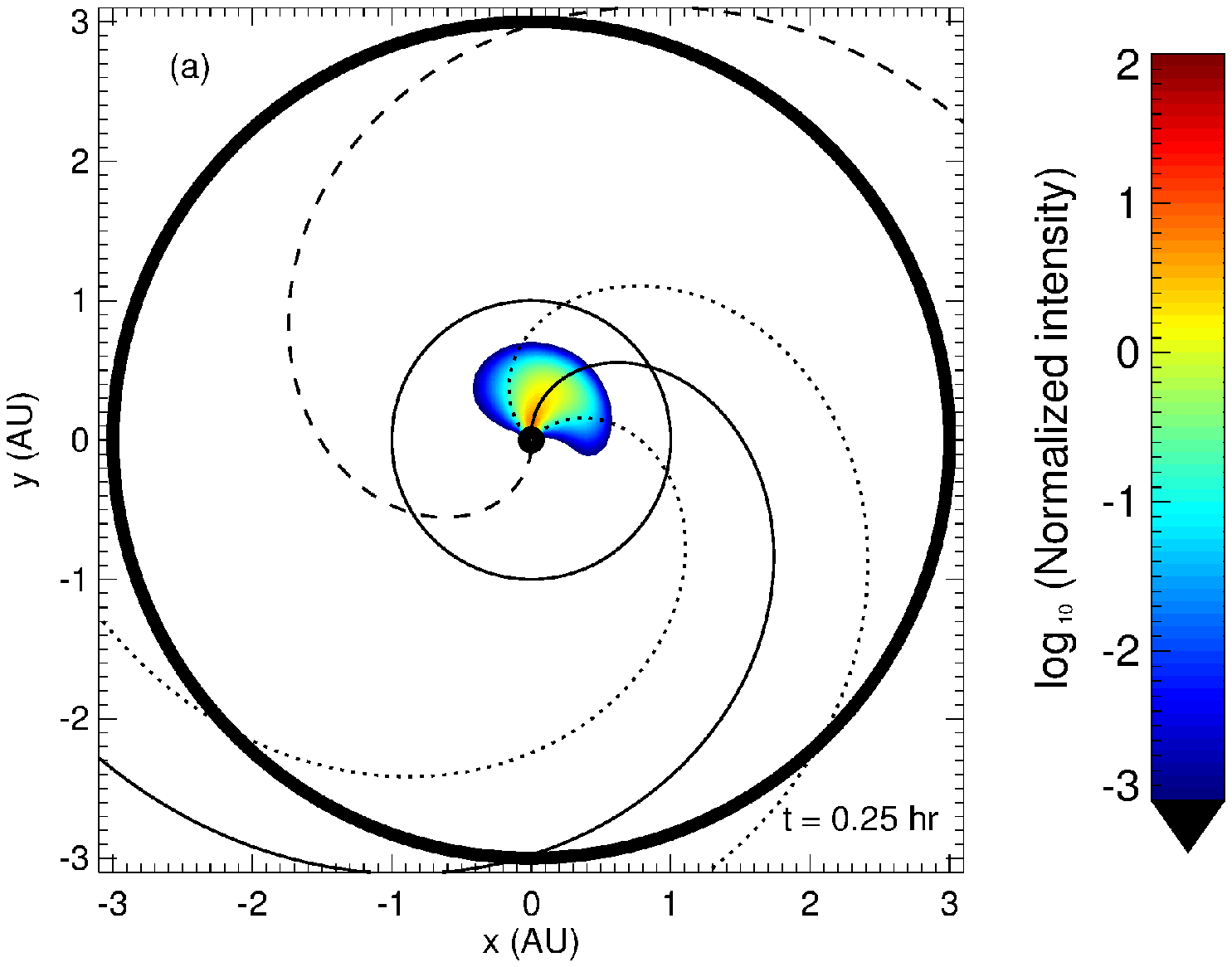} \includegraphics[height=60mm]{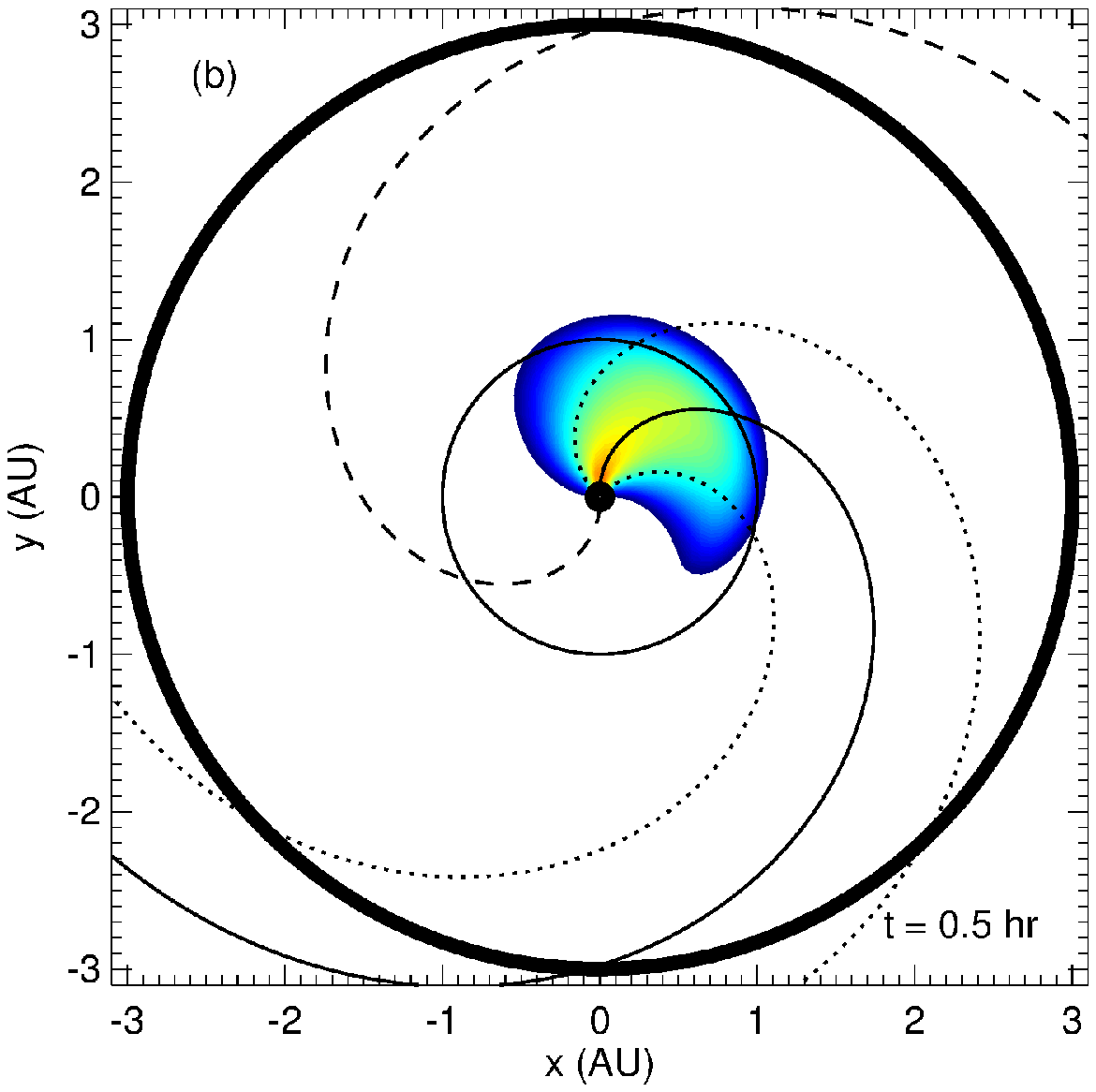} \\
\includegraphics[height=60mm]{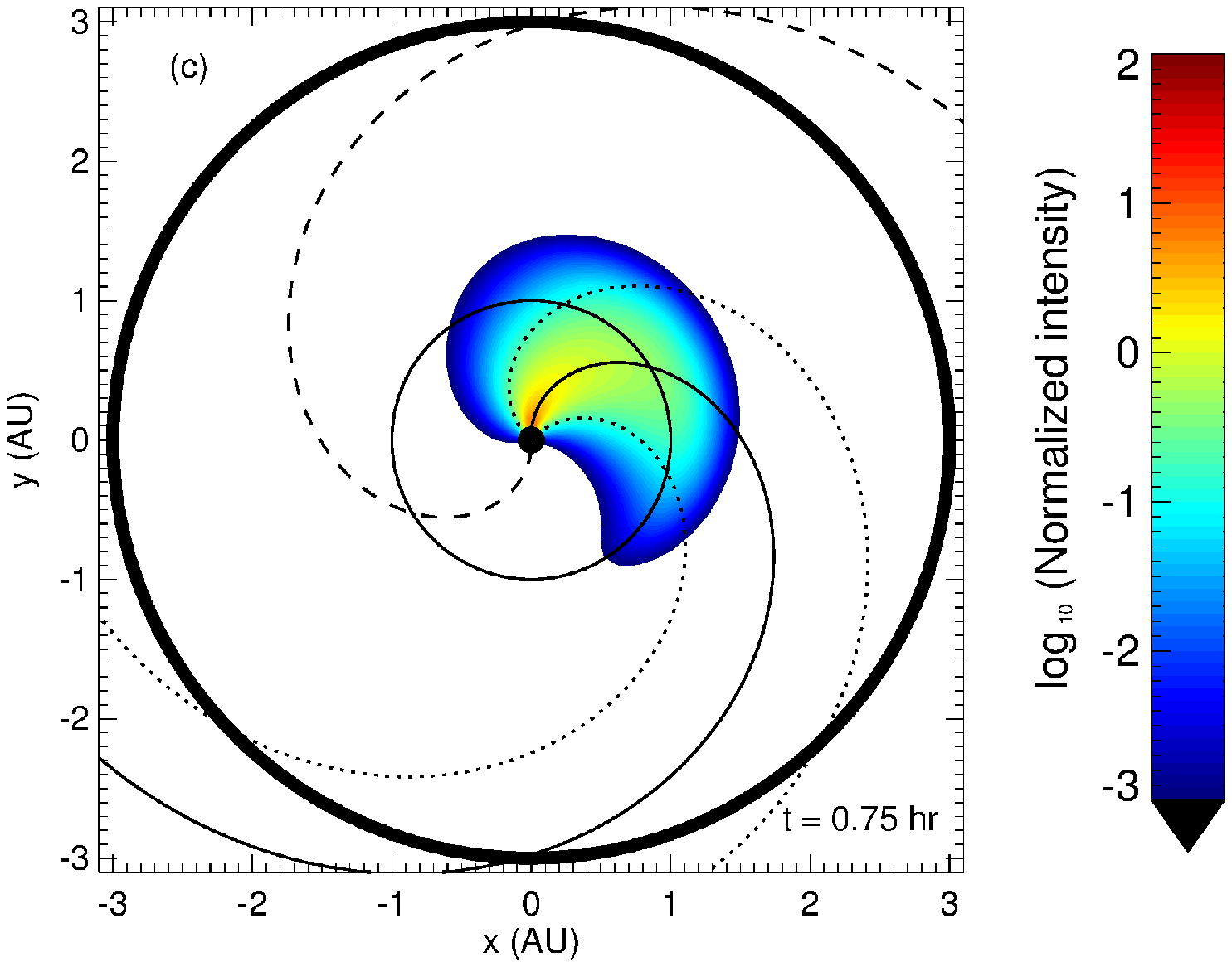} \includegraphics[height=60mm]{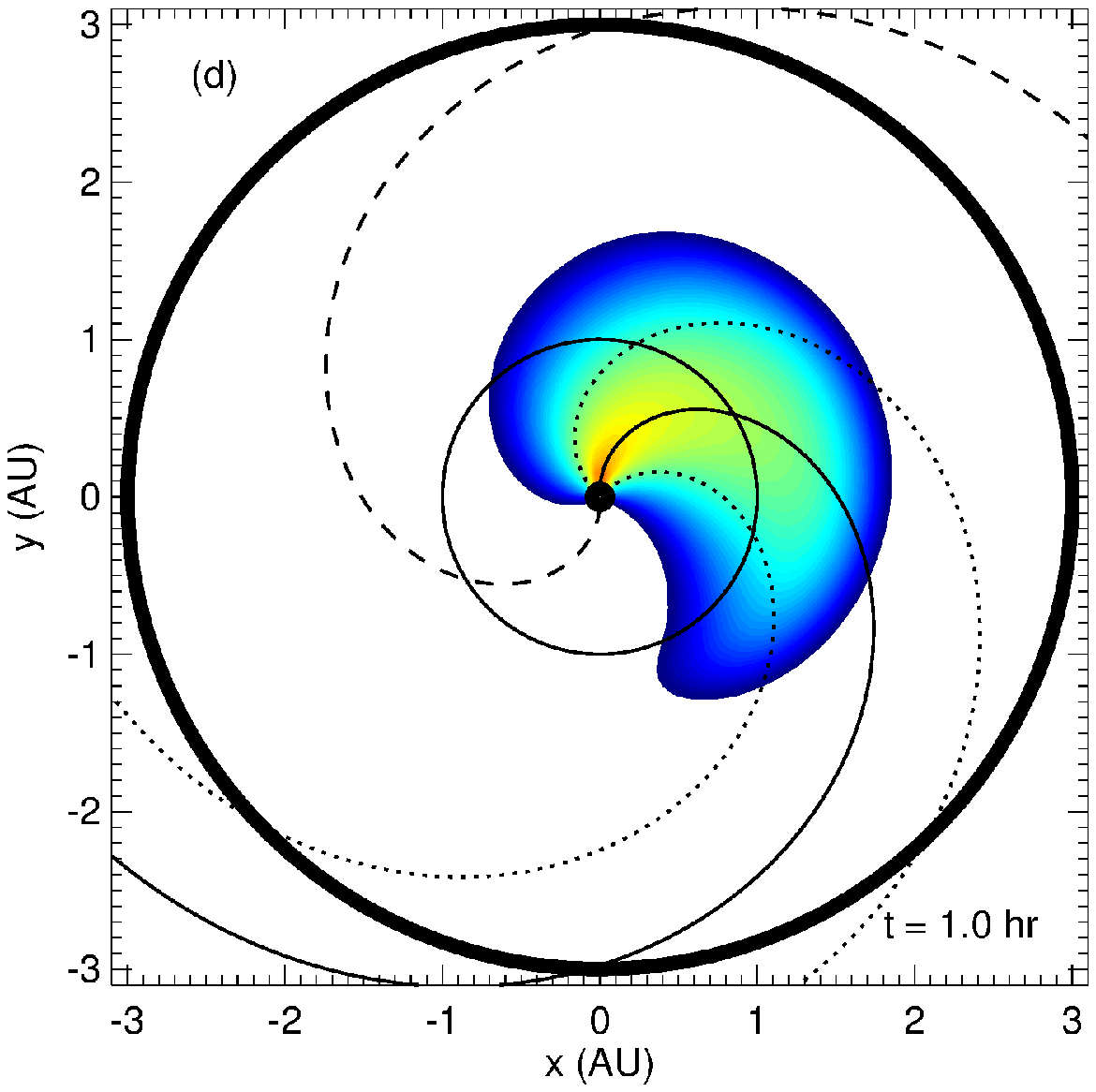} \\
\includegraphics[height=60mm]{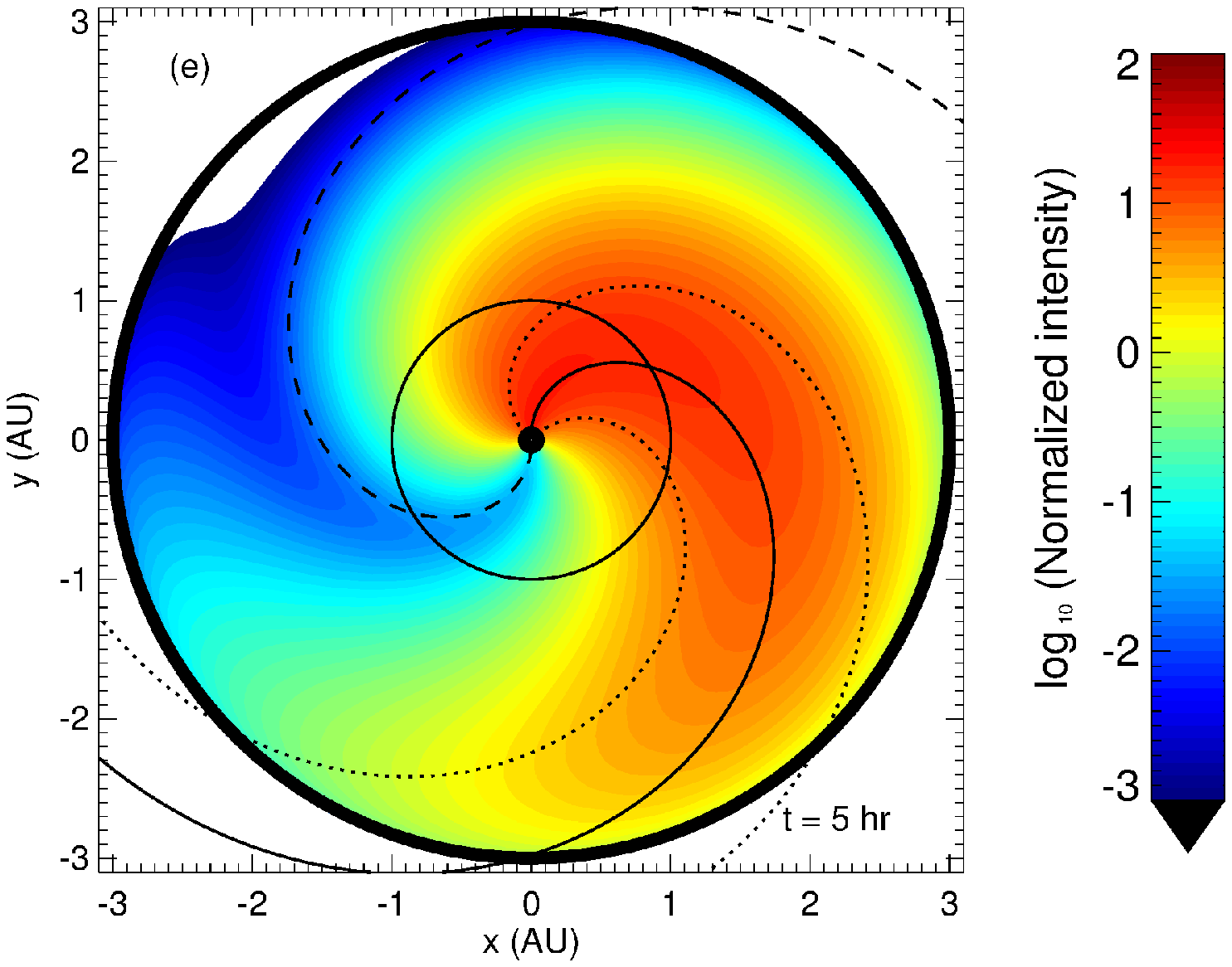} \includegraphics[height=60mm]{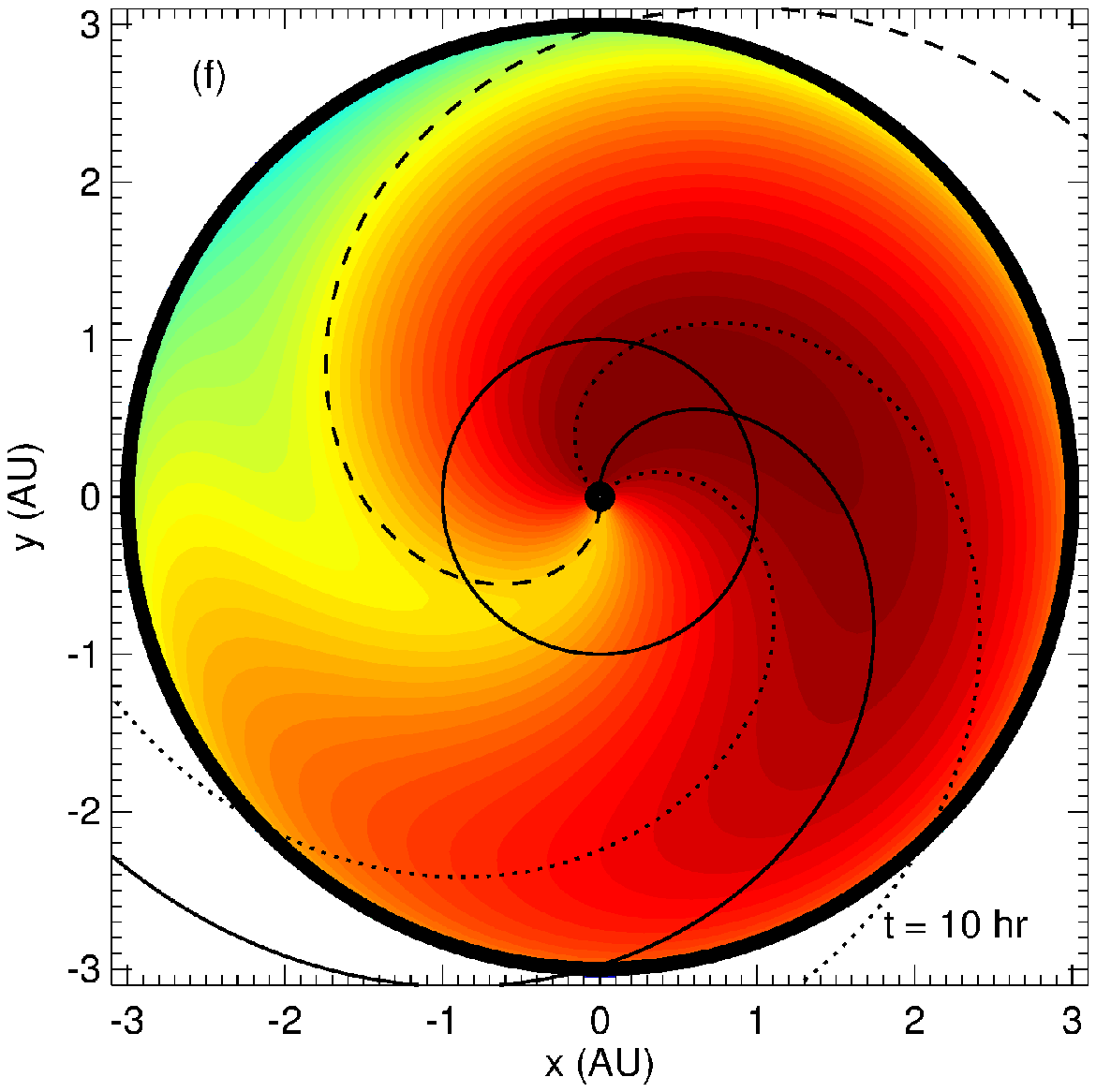} 
\caption{The temporal evolution of the normalized omni-directional intensity. For this simulation we used $\sigma = 5^{\circ}$ and $a=0.2$. See the text for additional details. \label{fig7}}
\end{figure*}

In Fig. \ref{fig7} we illustrate the temporal evolution of the normalized omni-directional intensity, for $\sigma=5^{\circ}$ and $a=0.2$, by showing contour plots at different times after particle injection. Panels (a) - (d) cover the first hour after injection while panels (e) and (f) are for later times. In the figure, the circles show the orbit of Earth and the model boundary, respectively, while different Parker magnetic field lines are added to guide the eye: the solid field line corresponds to the longitude where the injection function peaks (i.e. $\phi_0 = 90^{\circ}$ at the inner boundary), while the dashed lines are for $\phi_0 = 90 \pm 45^{\circ}$ showing the longitudinal extents of the injection function on this scale. In panel (a) we note that the SEP distribution has already spread significantly in longitude during the first 15 min of the simulation, extending beyond $90^{\circ}$ in longitude, and being much wider than the injected function. In order to examine the effectiveness of perpendicular diffusion, it is necessary to introduce the diffusive flux resulting from this process, which can be approximated as

\begin{equation}
\label{Eq:diff_flux}
\left| \mathcal{F}_{\perp}^{\mathrm{dif}} \right| \sim  D_{\perp} \left| \frac{\partial f}{\partial \phi} \right|.
\end{equation}

From this expression, it is clear that the effectiveness of perpendicular diffusion not only depends on the magnitude of the diffusion coefficient, but also on the associated gradient in the particle density. The latter quantity changes as a function of time, and, as we will show, so does the effectiveness of perpendicular diffusion. We injected SEPs over a very narrow region, resulting in large spatial gradients, so that, initially, perpendicular diffusion is very effective. In subsequent panels, (b) -- (f), the longitudinal spreading of the SEPs occurs at a much slower rate, i.e. the effectiveness of perpendicular diffusion decreases and the particle gradients get smaller. 

We therefore find that the rate of particle spreading, i.e. the effectiveness of perpendicular diffusion, is time-dependent and, due to our choice of a narrow injection function, the effectiveness of perpendicular diffusion decreases with time as the particle gradients tend to smear out due to a space-filling reservoir effect. This implies very efficient perpendicular diffusion early in a SEP event, even if $D_{\perp}$ is very small near the Sun. {Surprisingly, a similar conclusion was reached by \citet{Laitinenetal2013}, albeit on completely different grounds. These authors have shown that, early in a SEP event, particles follow wandering field-lines almost in a ballistic fashion (i.e. not decoupling from the field-lines), leading to very effective, but non-diffusive, transport. The effective perpendicular diffusion early in the event can also be} seen from Fig. \ref{fig5} where the onset times are relatively short over a wide range of longitudes: for $a=0.2$, a broad region, extending $\sim 120^{\circ}$ in longitude, has small onset times of less than 30 mins indicating that these particles must have diffused away from the source early in the SEP event. {However, these observations could also be produced by a significantly broader source region which we will address in the next section.}

\subsection{Broadness of the source region}

\begin{figure*}
\includegraphics[width=160mm]{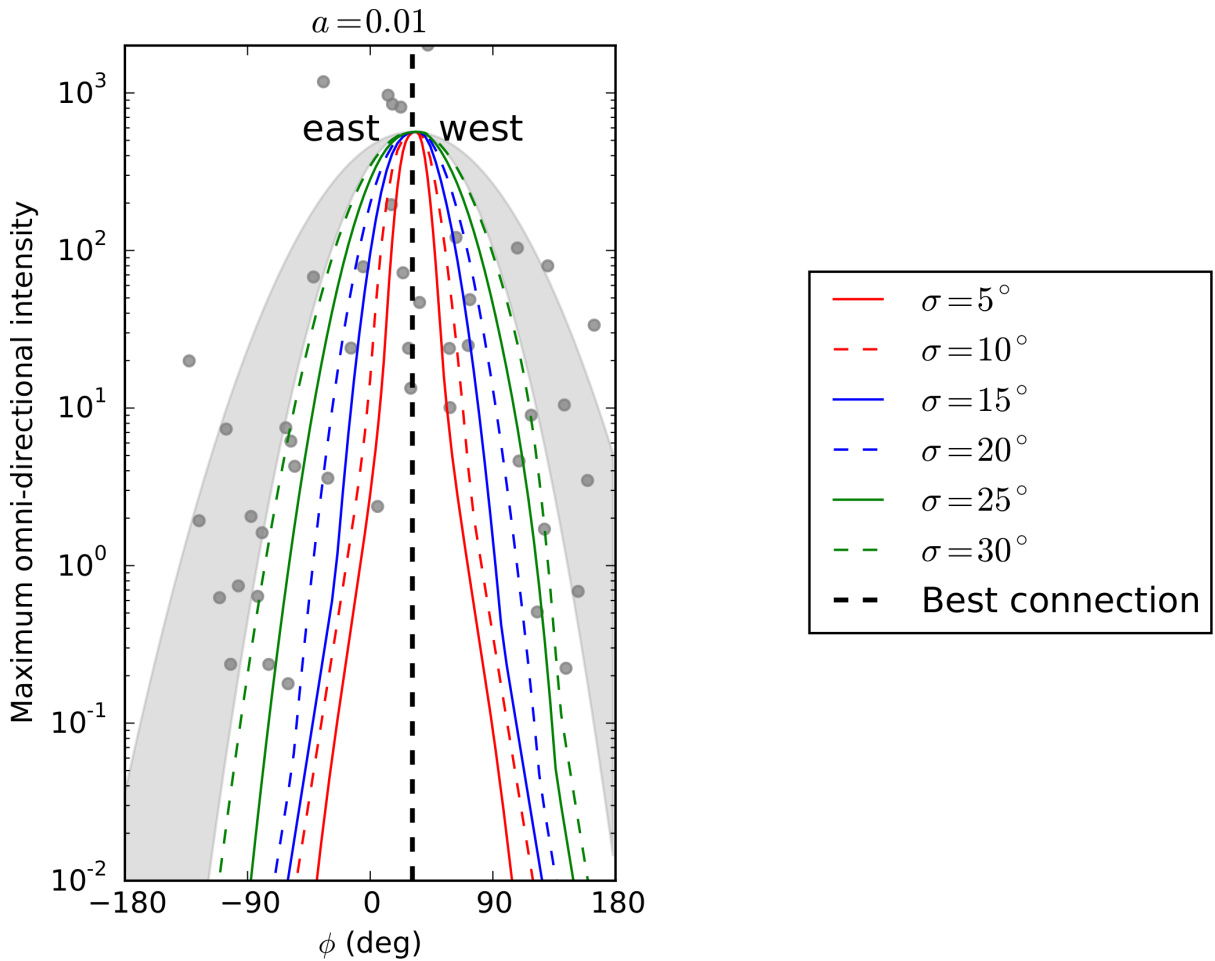}
\caption{Similar to Fig. \ref{fig4}, but now for $a=0.01$ held fixed and varying the injection broadness $\sigma$. \label{fig_source_1}}
\end{figure*}

\begin{figure*}
\includegraphics[width=160mm]{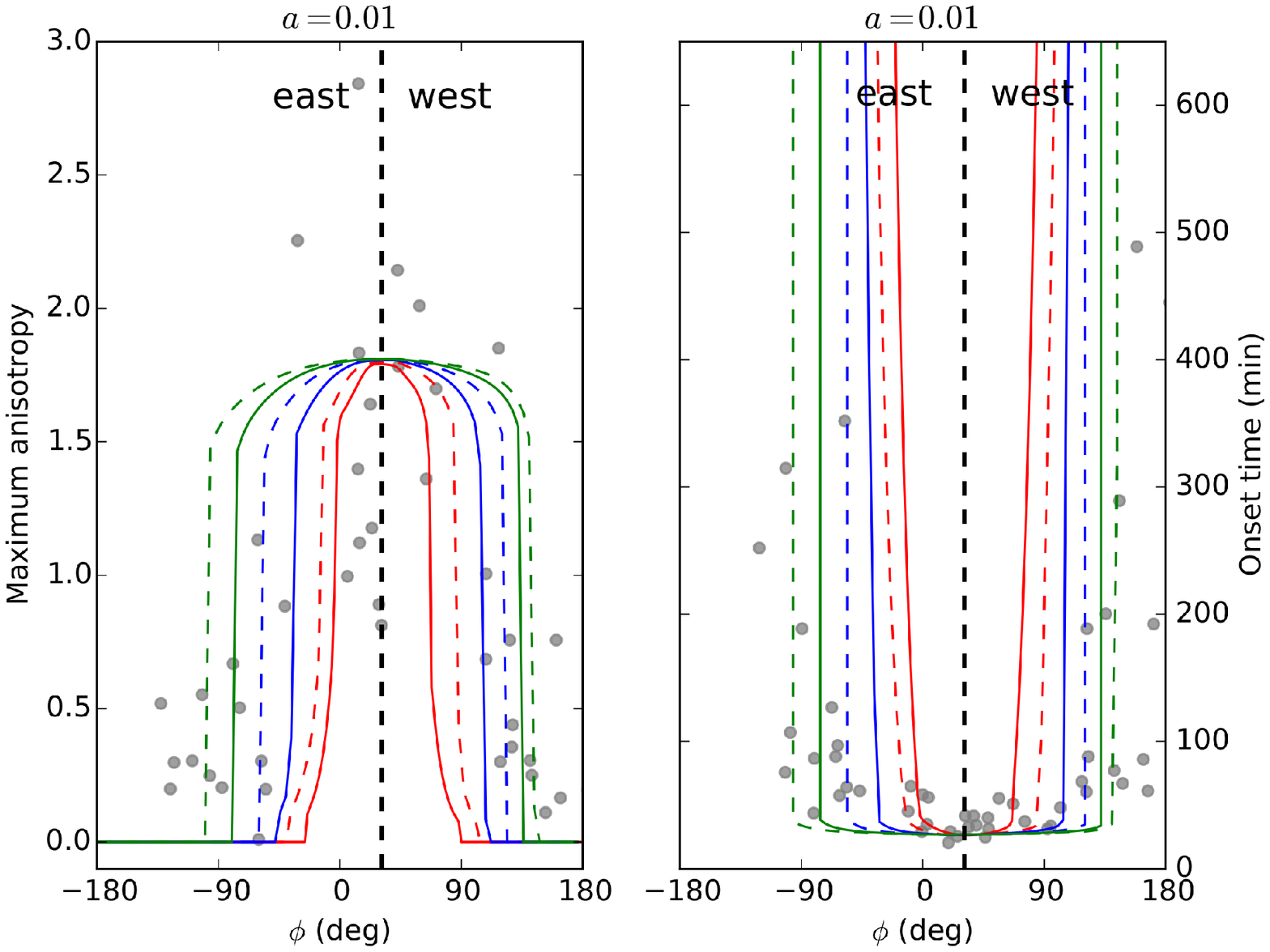}
\caption{Similar to Fig. \ref{fig5}, but now for $a=0.01$ held fixed and varying the injection broadness $\sigma$. \label{fig_source_2}}
\end{figure*}

{In the preceding sections, we have focused on the role of perpendicular diffusion by keeping the injection broadness, $\sigma=5^{\circ}$, constant in the model and varying the level of diffusion via the parameter $a$. Here, we keep $a$ fixed, and vary the {broadness of our Gaussian injection function}, $\sigma \in [ 5^{\circ}, 30^{\circ}  ]$, to examine the effect thereof on SEP propagation.}

{We start by assuming a value of $a=0.01$, which corresponds to very weak perpendicular diffusion. Such a scenario, where SEPs are injected over a wide range of longitudes, and suffer little or no perpendicular diffusion has been proposed by e.g. \citet{cliveretal1995} and \citet{remaesetal1996} as an explanation for the observed broadness of the SEP distribution. The calculated maximum omni-directional intensity, for this scenario, is shown in Fig. \ref{fig_source_1}, as a function of longitude at Earth's orbit, for different choices of $\sigma$. As expected, when a broader source function is introduced, the resulting distribution at Earth broadens. However, it should be noted that, even for an injection function with a broadness of $\sigma = 30^{\circ}$, the simulation results are still well below the observed broadness of the widespread events (the shaded gray region). }

{In Fig. \ref{fig_source_2}, which is similar to Fig. \ref{fig5}, we calculate the maximum anisotropy and the onset times for the scenario where $a=0.01$ is held fixed, but the broadness of the injection region is varied. Again, the results are mostly in the expected form. The maximum anisotropy values are either high (around a value of 1.5) when one is magnetically connected to the source, or essentially zero when not magnetically connected to the source. Similarly, the onset times are either very short (close to 20 mins) when magnetically connected, or exceedingly long (the simulations were run for 75 model hrs and the calculated intensity was still significantly below the assumed background level). These results illustrate the ``binary" nature of assuming a broad source and no other perpendicular transport mechanisms: you are either connected to the source (high anisotropy values and short onset times) or not (no event is observed). These simulation results are difficult to reconcile with the observations which show a clear distribution for both of these quantities.  }

\begin{figure*}
\includegraphics[width=160mm]{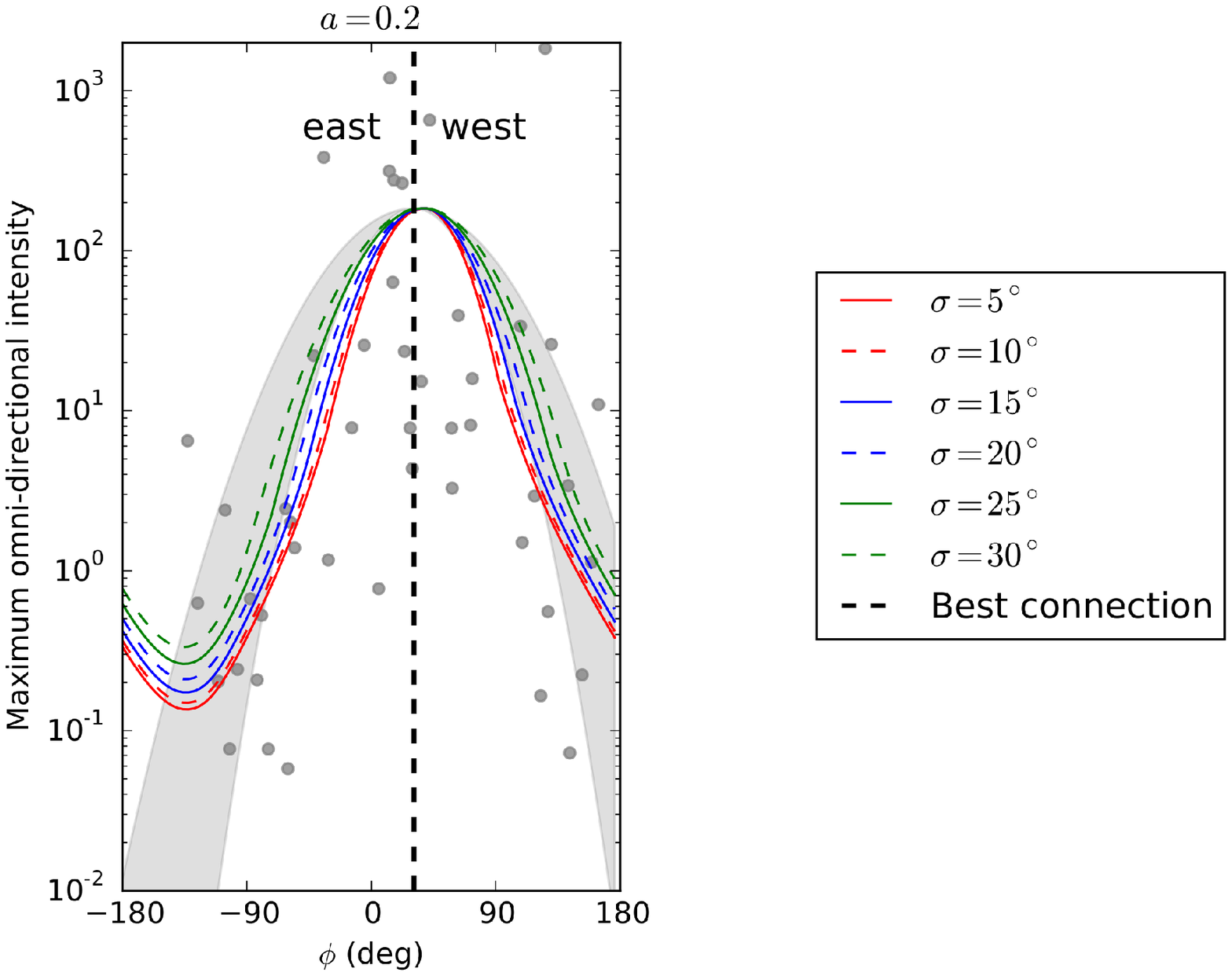}
\caption{Similar to Fig. \ref{fig_source_1}, but now for $a=0.2$. \label{fig8}}
\end{figure*}

\begin{figure*}
\includegraphics[width=160mm]{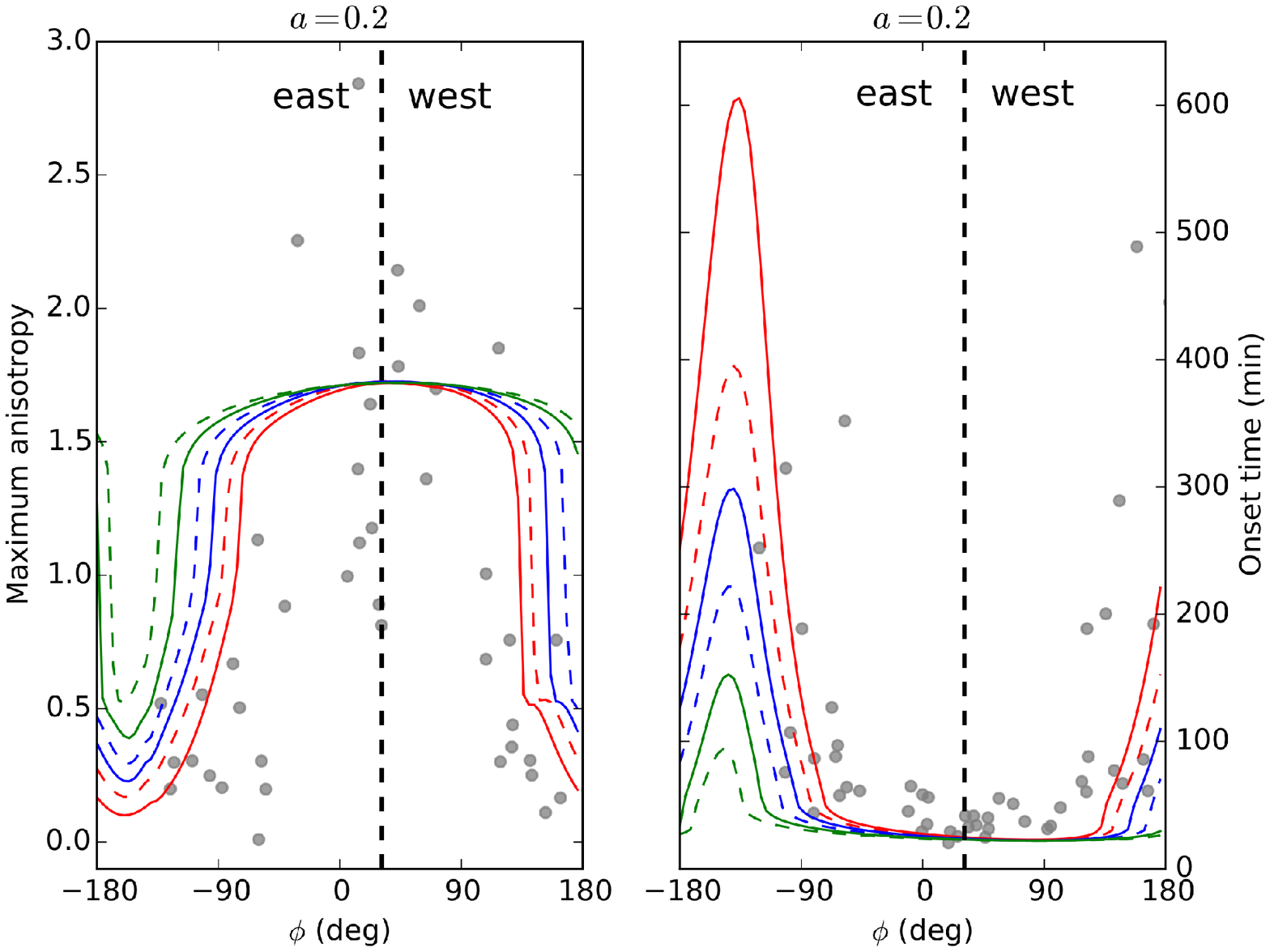}
\caption{Similar to Fig. \ref{fig_source_2}, but now for $a=0.2$. \label{fig9}}
\end{figure*}

{For the next simulations, we include perpendicular diffusion in the model and choose a value of $a=0.2$ as this results in both reasonable values of $\lambda_{\perp}$, and also a fair comparison with observations in preceding figures (see Figs. \ref{fig4} and \ref{fig5}). The calculated maximum omni-directional intensity is shown in Fig. \ref{fig8}, as a function of longitude at Earth's orbit, for different choices of $\sigma$. Rather unexpectedly, we note that the intensities are not very sensitive to the broadness of the injection function. This can, however, be explained by again looking at Eq. \ref{Eq:diff_flux}: for a narrow injection, large particle gradients are present, resulting in very effective perpendicular diffusion. By now increasing $\sigma$, we are actually decreasing the effectiveness of the diffusion process. The complex and non-linear interplay between the magnitude of the perpendicular diffusion coefficient and the intensity gradient driving perpendicular diffusion is therefore shown to be important and must be kept in mind when interpreting these model results.}

{In Fig. \ref{fig9}, which is similar to Fig. \ref{fig5}, we calculate the maximum anisotropy and the onset times for the scenario where $a=0.2$ is held fixed, but the broadness of the injection region is varied. As expected, a broader injection region leads to a broader spreading of higher anisotropy values and shorter onset times away from best magnetic connection. However, when comparing these to the results shown in Figs. \ref{fig5} and \ref{fig_source_2}, we note that the model results are much more sensitive to the choice of $a$ (the level of perpendicular diffusion) than to $\sigma$ (the broadness of the injection region). }

\subsection{Different forms of the source region}

\begin{figure}
\includegraphics[width=80mm]{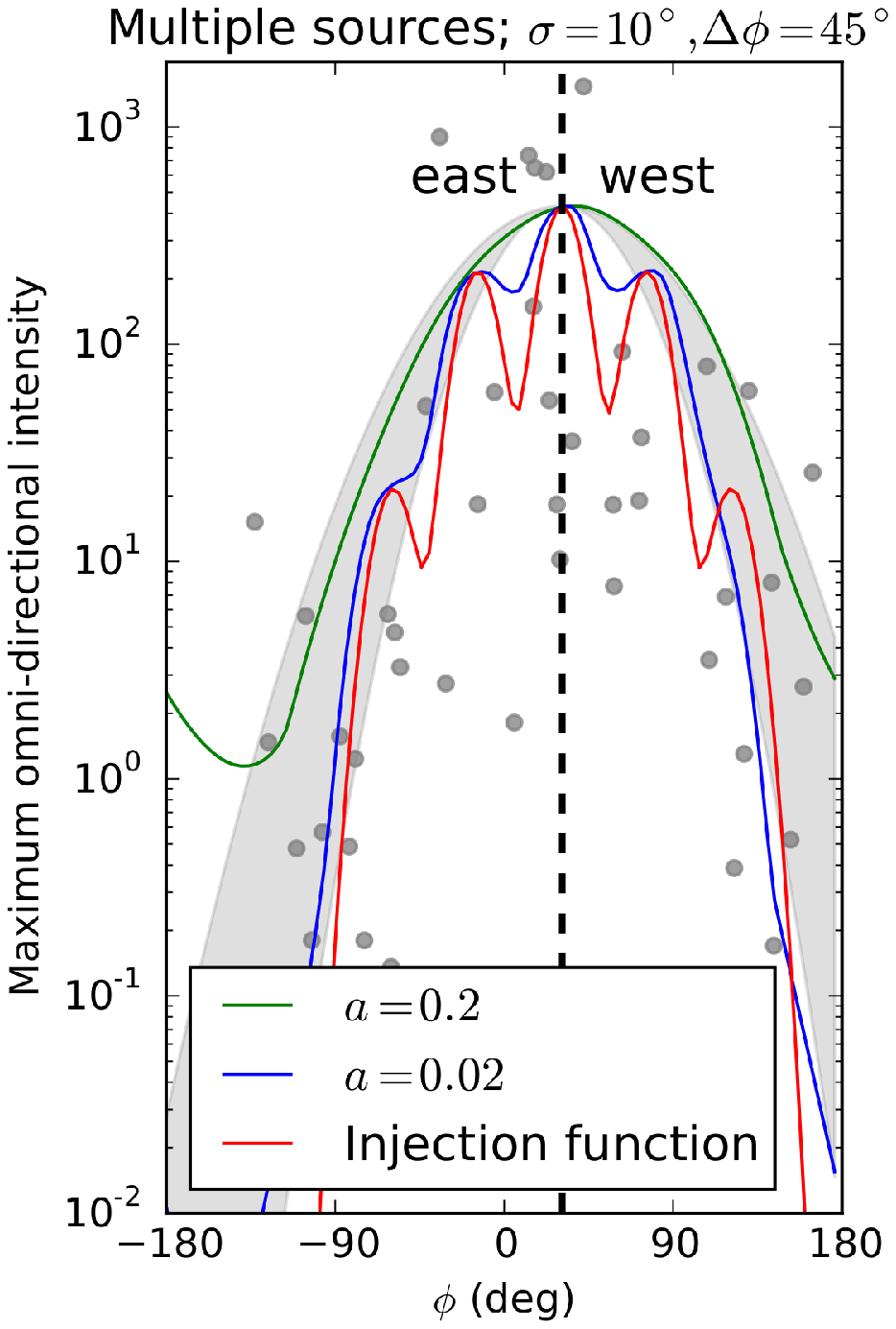}
\caption{Similar to Fig. \ref{fig4}, but here we have chosen a non-Gaussian form for the injection function (solid red line) and calculated the intensity at Earth for two different levels of perpendicular diffusion (the values of $a=0.02$ and $a=0.2$ correspond to the blue and green model solutions respectively). \label{fig10}}
\end{figure}

In this section we examine what effect the form of the injection function may have on particle intensities. This is motivated by the observations of \citet{klassenetal2016}, suggesting that during some SEP events, the SEPs may be injected into different flux tubes resulting in a longitudinal distribution having multiple, finger-like, peaks (see their Fig. 16). We introduce such a complex injection function in the model and show the results in Fig. \ref{fig10}. Here, the injection function (red line) consists of a combination of 5 Gaussian functions (each with $\sigma = 10^{\circ}$), spread $45^{\circ}$ apart. If we assume weak levels of perpendicular diffusion ($a=0.02$, blue curve), the complexity of the injection function is preserved at Earth, but for ``normal" levels of diffusion ($a=0.2$, green curve), the complexity is simply smeared out and a rather featureless distribution is obtained. {Therefore, in order to reproduce a SEP event at Earth showing multiple finger-like peaks, we not only need injection of particles into different flux tubes, as suggested by \citet{klassenetal2016}, but we would also somehow need to considerably reduce the level of perpendicular diffusion from ``normal" levels. }

\subsection{Multiple injections from the same active region}
\label{Sec:multiple}

\begin{figure*}
\includegraphics[width=120mm]{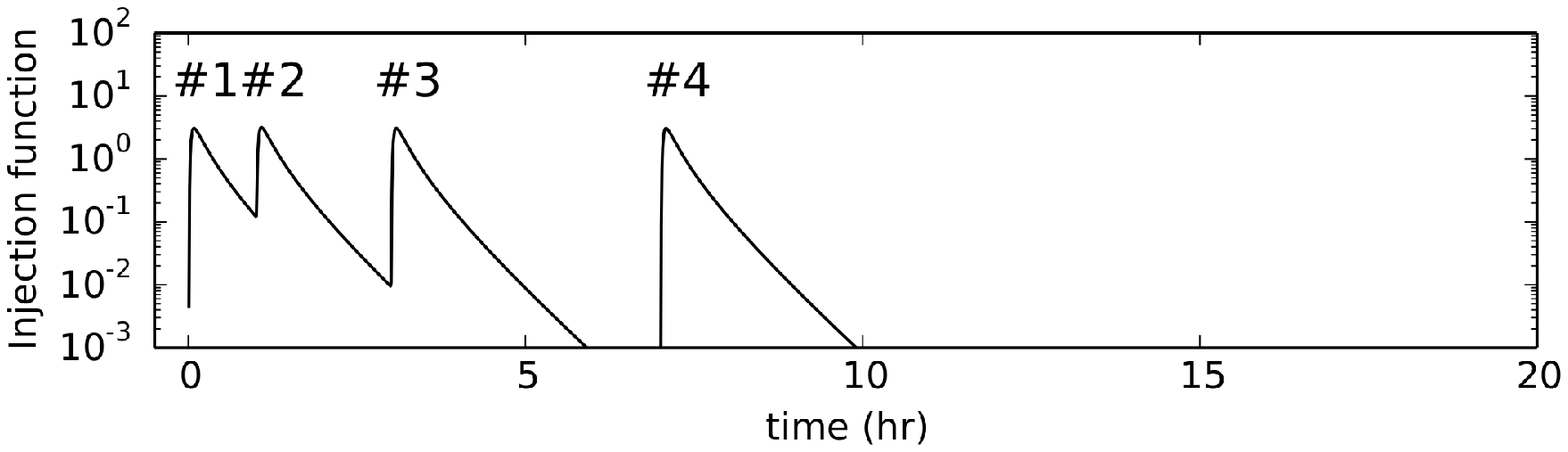}
\caption{The four injection functions introduced in the model at the inner boundary. \label{fig:multiple_injections}}
\end{figure*}

In a recent paper, \citet{aguedalario} investigated a SEP event which was simultaneously observed by the {\it Helios-1} and {\it IMP-8} spacecraft, and showed {multiple SEP events} within a $\sim 20$ hr period. Interestingly, they showed that the amplitude of the time-dependent SEP flux varied significantly at {\it Helios-1} (which is, of course, much nearer to the Sun), while a more gradual and isotropic distribution was recorded at Earth. Interpreting this in terms of only pitch-angle scattering along a single magnetic field lines is difficult \citep[see the modeling results and discussion by][]{aguedalario}, and hence, we re-investigate such a scenario in our present model where perpendicular diffusion is included. Instead of a single injection at $t=0$, as was done up to now in the model, we repeatedly inject four isotropic distributions at the inner boundary at $t=0,1,3,7$ hrs using the same profile as given in Eq. \ref{Eq:reid_injection}, except that $\tau_e = 0.5$ hrs. The temporal profile of this new combined injection profile is shown in Fig. \ref{fig:multiple_injections} at $\phi = \phi_0 = \pi/2$ and $r=r_0$. For all injections, we assume $\sigma = 5^{\circ}$ and keep $C$, the magnitude of the injection function, constant, but could, in future, vary all of these parameters independently. For the subsequent simulations we keep $a=0.2$ fixed.

\begin{figure*}
\includegraphics[width=120mm]{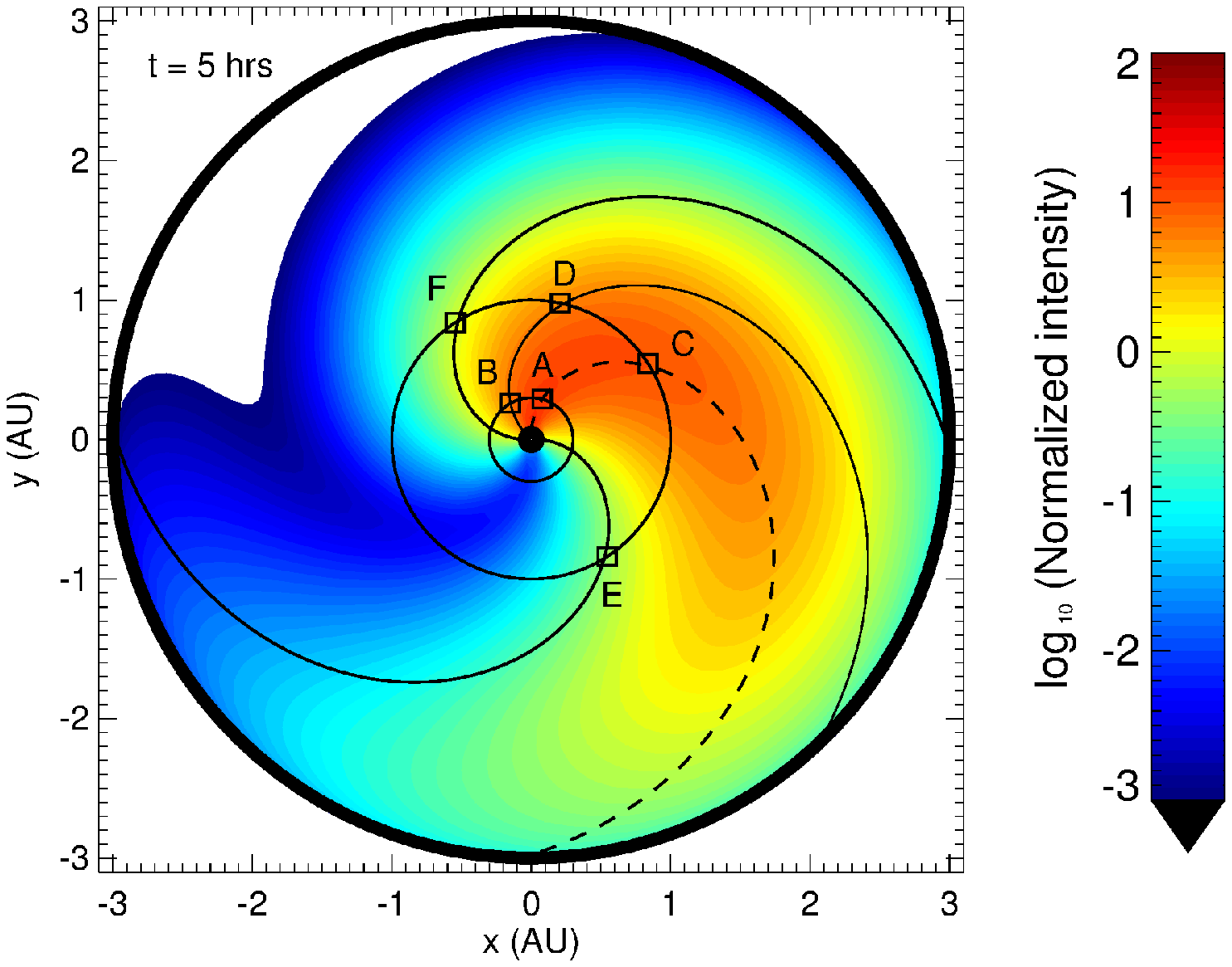}
\caption{The resulting SEP intensities at $t=5$ hrs when multiple sources are injected at different times. Also shown, as the small squares, are the position of six virtual spacecraft at which we will calculate the SEP intensity as a function of time. \label{fig:multiple_injections_contour}}
\end{figure*}

\begin{figure*}
\includegraphics[width=120mm]{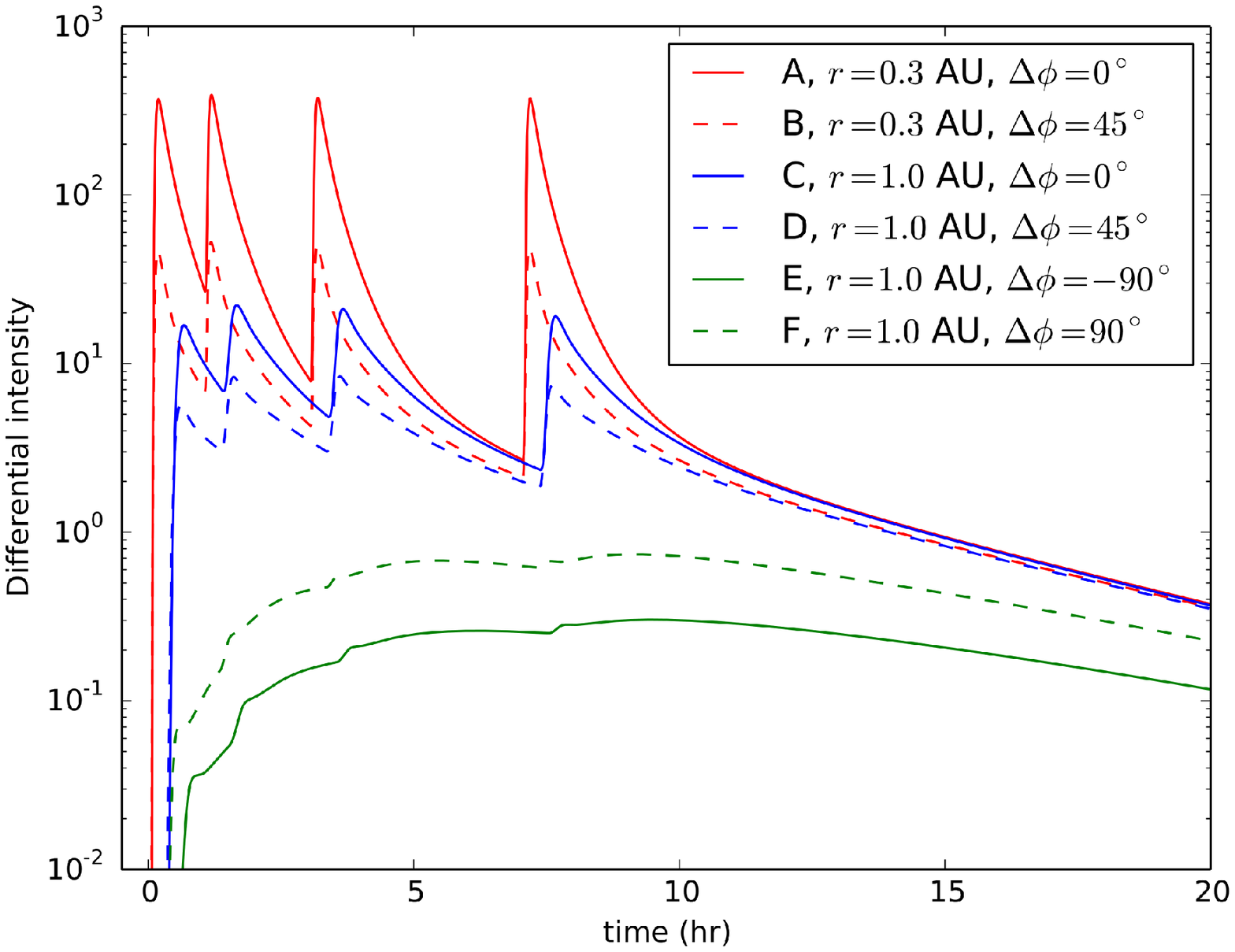}
\caption{The resulting omni-directional intensity, as a function of time, at the different spacecraft poisitions indicated on Fig. \ref{fig:multiple_injections_contour}. \label{fig:multiple_injections_intensity}}
\end{figure*}

Fig. \ref{fig:multiple_injections_contour}, which is similar to Fig. \ref{fig7}, shows the modeled distribution at $t=5$ hrs. In the figure, we have also included the approximate orbits of the {\it Helios-1} ($\sim 0.3$ AU) and {\it IMP-8} spacecraft (Earth, $\sim 1$ AU) and the position of six virtual spacecraft at which we will track the temporal evolution of the SEP distribution (these are indicated by the labelled squares). A and C are magnetically connected to the source function's maximum, while B and D are $\Delta \phi = 45^{\circ}$ away (towards the west), but still relatively well connected. E and F are definitely not magnetically connected to the source, $\Delta \phi = \pm 90^{\circ}$, and SEPs reaching these positions must therefore experience significant perpendicular diffusion. The calculated omni-directional intensities at these spacecraft positions are shown in Fig. \ref{fig:multiple_injections_intensity} as function of time. For spacecraft A and B, both being within $\sim 0.3$ AU of the source, and relatively well connected to it, the intensity profile follows that of the injection function: four well-defined peaks are noted. This behavior is expected as $\lambda_{||} \gg L$ close to the Sun (see again Fig \ref{fig3}), and these particles propagate almost ballistically from the source to the {\it Helios-1} orbit. At the positions of C and D (at Earth, but still relatively well connected to the SEP source), we again note four well-defined peaks, albeit with a smaller amplitude. Pitch-angle scattering between $\sim 0.3 - 1$ AU therefore seems to be able to somewhat reduce the amplitudes of the different peaks, but cannot produce {a single gradual SEP event} at Earth \citep[a similar conclusion was made by][]{aguedalario}. Lastly, at positions E and F (again at Earth's orbit, not magnetically connected to the source), we note an almost featureless gradual SEP event. We therefore conclude that a combination of pitch-angle scattering and perpendicular diffusion is able to reproduce {the \citet{kallenrode} observations, as discussed by \citet{aguedalario},} by smearing-out the temporal profile of the injection function very effectively.

\begin{figure*}
\includegraphics[width=120mm]{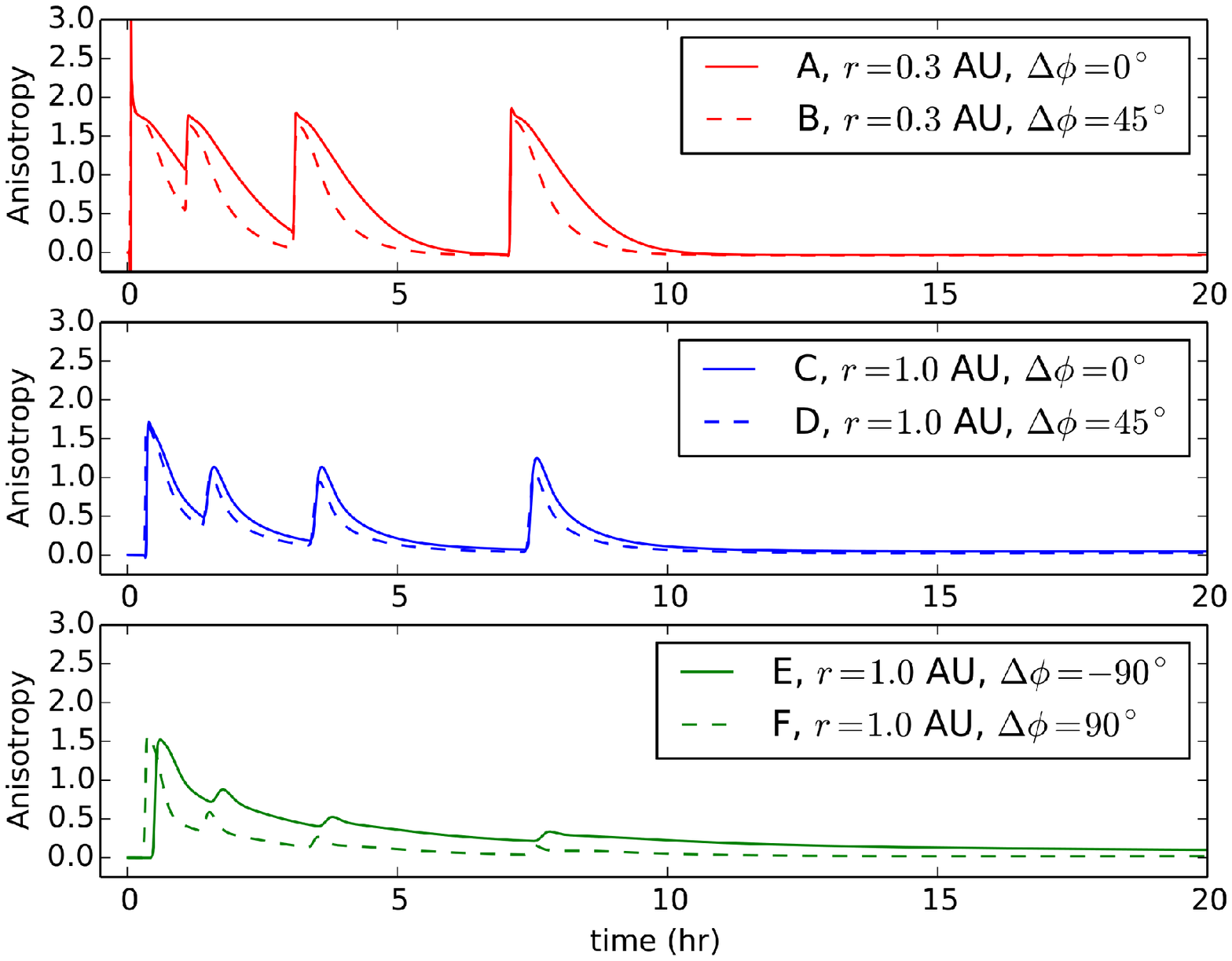}
\caption{Similar to Fig. \ref{fig:multiple_injections_intensity}, but now showing the calculated anisotropy. \label{fig:multiple_injections_anisotropies}}
\end{figure*}

The corresponding anisotropies at the positions of the six virtual spacecraft are shown in Fig. \ref{fig:multiple_injections_anisotropies}. As expected, the anisotropies are very large close to the Sun (A and B), decrease to moderate values at Earth's orbit when SEPs reach these positions mainly by parallel transport (C and D), and are very small at large radial distances where perpendicular diffusion played an important transport role (E and F{, this is true for injections $\#2-\#4$, while the anisotropy corresponding to the first injection events is still large at all positions}). 

The behavior of both the modeled SEP intensity and anisotropy is in {general agreement with the results presented by \citet{aguedalario}: when SEPs} propagate mainly through perpendicular diffusion to reach Earth, their resulting intensities are relatively independent of the temporal form of the injection function and such an observer (e.g. being at either point E or F) would register only a gradual, and isotropic, SEP event. This is related to the reservoir, or space-filling, effect described by e.g. \citet{hewang2015}, where perpendicular diffusion results in a nearly constant SEP flux throughout the inner heliosphere at late times.

\section{Discussion}

In this paper we have tried, as far as possible, to model SEP electron propagation by implementing theoretically derived transport coefficients. For $D_{\mu\mu}$ we implemented a variation of the well-known plasma wave model for slab turbulence and for $D_{\perp}$ we assumed that the 2D fluctuations will scatter the electrons via the FLRW process. Both of these coefficients are not dependent on any {\it ad-hoc} free parameters, but rather related to the underlying turbulence in the magnetic field which, in turn, can be determined (or at least, constrained) by further observations. {Our choices of these fundamental turbulence quantities are in line with those used by e.g. \citet{EB2013} for radial distances of 0.3 AU and larger. The validity of these assumptions closer to the Sun, which is of extreme importance for SEP transport, can only be confirmed by future space missions, most notably the upcoming {\it Solar Probe Plus} mission.} The only remaining free parameters in the model are $\sigma$ (the broadness of the injection function) and $a$ (related to the probability of SEPs following turbulent field lines), and these are varied in the model.

We find that the resulting coefficients, especially the radial dependence of $\lambda_{||}$ and $\lambda_{\perp}$, are probably much more complex than previously thought and would be difficult to described phenomenologically. Of special interest is the dependence of $\lambda_{\perp}$ which decreases towards the Sun leading to a ratio $\lambda_{\perp}/\lambda_{||}$ that varies between $\lambda_{\perp}/\lambda_{||} \sim 0.001 - 0.03$ in the model when choosing a reasonable estimate of $a=1/10$. Although this would imply very weak perpendicular diffusion close to the Sun (i.e. early in a SEP event), we however still find effective diffusion in this regime due to the large particle gradients present here (see again Eq. \ref{Eq:diff_flux}). We emphasize that the effectiveness of perpendicular diffusion is determined by both the magnitude of the diffusion coefficient, and the magnitude of the particle gradient which drives perpendicular diffusion. The effectiveness of perpendicular diffusion in our model thus changes time-dependently, being extremely effective early in the event. This behavior is especially evident when looking at the calculated onset times: centered at best magnetic connection, extending almost $\sim 120^{\circ}$ in longitude, we find short onset-times for narrow injection scenarios, indicating that these particles propagated very effectively in longitude early in the SEP event close to the Sun. The onset times increase further away from best connection and, as perpendicular diffusion becomes an increasingly slower process, it takes increasingly longer to fill the entire computational regime with SEPs late in the event. Generally, we also find that longer onset times correspond to smaller anisotropies, confirming the general belief that longer propagation times lead to more scattering and a more isotropic SEP distribution.

The model solutions are, as expected, very sensitive to the choice of $a$, as illustrated in Fig. \ref{fig4}. We however believe that the extreme case of $a=1$ is unreasonable and a rough comparison with observations suggests that a value of $a=0.2$ seems plausible and would be a reasonable estimate of this quantity.  Our estimate of $a$, unsurprisingly, differs from the generally used value of $a=1/\sqrt{3} \sim 0.6$ used mainly for protons \citep[][]{Matthaeusetal2003}. Additional numerical simulations of the transport coefficients, focusing especially on electrons (small gyro-radii), such as these presented by \citet{hausseinetal2016}, are needed to further constrain this quantity and its possible energy dependence. 

For a ``reasonable" choice of $a=0.2$, our results are very insensitive to the form (longitudinal dependence) of the injection function. Different choices of $\sigma$, which determines the broadness of the injected function, seemingly lead to the same longitudinal distribution at Earth. This is, however, again explainable through the effectiveness of perpendicular diffusion (see again Eq. \ref{Eq:diff_flux}). Moreover, by assuming more exotic forms of the injection function, as in Fig. \ref{fig10}, perpendicular diffusion seems to smear away any small-scale features of the injected distribution, leading to a rather featureless Gaussian-like distribution at Earth. This is however based on the assumption of perpendicular diffusion in a regular large-scale background Parker field and would not be valid if diffusion barriers or magnetic structures would impede the diffusion process \citep[see, for instance,][]{straussetal2016}.

By examining our calculated transport coefficients (see again Fig. \ref{fig3}), we note that, near the Sun at the approximate position of the {\it Helios-1} spacecraft, $\lambda_{||} \gg L$, indicating that SEPs would propagate almost ballistically from their source to such a spacecraft. This is consistent with the findings of \citet{kunow}. 

Motivated by the observations presented by \citet{aguedalario}, we investigated the effect of multiple injections of SEPs from the same active region in Section \ref{Sec:multiple}. Our results show that a SEP event, having a very complex temporal profile near the Sun, can be observed to be rather featureless, and nearly isotropic, at Earth's orbit when SEPs are transported to the latter position mainly via perpendicular diffusion. 

Our results suggest that perpendicular diffusion remains a viable process to explain the large longitudinal spread of SEP electrons. {We expand on previous work that has included perpendicular diffusion \citep[e.g. those of][amongst others]{zhangetal2009,wolfgang,drogeteal2014,he2015}, by included a theoretical treatment of the transport parameters rather than adopting {\it ad-hoc} expressions.} Not only are we able to reproduce the observed spread of particle intensities, but also the longitudinal dependence of the onset times and maximum anisotropies. Although the contribution of other processes to the longitudinal spread of SEPs cannot be neglected, the important role of perpendicular diffusion should be clear. {It should be kept in mind that some authors favour a scenario where SEPs are injected over a wide range of longitudes and suffer little or no perpendicular diffusion \citep[e.g.][]{cliveretal1995,remaesetal1996,kahler2016}, in contrast to the findings presented here which largely excludes this scenario.} However, some fundamental questions remain unanswered, such as what perpendicular diffusion process is actually dominant and which energies? We have assumed that the FLRW process is a good approximation for the diffusion of low {energy electrons with energies $\sim 100$ keV}, but this is unlikely to be the case for high energy electrons or SEP protons of any energy. {Therefore, before extending our present model to simulate proton or high energy electron transport, we would need to include co-rotation, drifts and energy losses in the model, while a theoretical $D_{\perp}$, based on a non-linear theory (which is required for particles with large Larmor radii), and containing a $\mu$-dependence, still does not exist.} {Moreover, the connection between the FLRW coefficient and so-called field line meandering \citep[as introduced by][]{Laitinenetal2016} remains vague. In the latter process, particles follow wandering (meandering) field-lines without de-coupling from them, leading to very effective, but non-diffusive, cross-field transport. Recently, \citet{Laitinendalla2016} have presented simulations that suggest that field-line meandering may only be the initial phase of the cross-field diffusion process, where, at later times when the particles decouple from their initial field-lines, the FLRW diffusion process is recovered. It is unclear whether electron cross-field transport between the Sun and Earth is governed by the initial or late phase of diffusion. This is connected} to the assumption that perpendicular transport can be described as a diffusive process close to the Sun, as we have done here, or if we need to model it through deterministic means \citep[see again][]{Laitinenetal2016}, or more fundamentally, by following particle orbits \citep[see the model of][]{Ablassmayeretal}. These questions cannot be easily answered and would need much more modeling and observational input. Here, future space missions (such as {\it Solar Probe Plus} and {\it Solar Orbiter}) may play an important role: by simultaneously observing a SEP event, a constellation of spacecraft, at different longitudes and radial positions, may give us an idea of how the broadness of the SEP intensity varies, in addition to other observables. These spacecraft may also help constrain the levels and forms of the turbulence spectra which are the main input parameters to this and other transport models.

\acknowledgments

{This work is based on research supported in part by the National Research Foundation (NRF) of South Africa (grant no. 106049). Opinions expressed and conclusions arrived at are those of the authors and are not necessarily to be attributed to the NRF. N.D. was supported under grant 50OC1302 by the Federal Ministry of Economics and Technology on the basis of a decision by the German Bundestag.} {This work was carried out within the framework of the bilateral BMBF-NRF-project `Astrohel' (01DG15009) funded by the Bundesministerium f\"ur Bildung und Forschung and the NRF. The responsibility of the contents of this work is with the authors.}



\end{document}